\documentclass[a4paper,11pt]{article}
\usepackage[english]{babel}    
\usepackage{jcappub} % for details on the use of the package, please see the JINST-author-manual
\usepackage{lineno}
\usepackage{graphicx}
\usepackage{mathtools}
\usepackage{xcolor}
\usepackage{babel}
\usepackage{hyperref}
\hypersetup{
	colorlinks = true,
	linkcolor = {blue}
}
\newsavebox{\arrangebox}

\usepackage{amsmath}
\usepackage{multirow}

\usepackage{colortbl}
\usepackage{adjustbox} % for resizing the table

\arxivnumber{2503.00532} % Only if you have one
\title{\boldmath Performance and radiation damage mitigation strategy for silicon photomultipliers on LEO space missions}
\author[a]{L. Burmistrov}
\author[a]{, S. Davarpanah$^*$}
\author[a]{, M. Heller}
\author[a]{, T. Montaruli}
\author[a]{,\\ C. Trimarelli$^*$}

\author[b, c]{, R. Aloisio}
\author[b, c]{, F.C.T.~Barbato}
\author[b, c]{, I. De Mitri}
\author[b, c]{,\\ A. Di Giovanni}
\author[b, c]{, G. Fontanella}
\author[b, c]{, P. Savina}
\author[a]{, C. T\"onnis}
\author[d]{,\\E. Moretti}
\author[d]{, M. Ruzzarin}
\author[e]{, J. Swakoń}
\author[e]{, D. Wróbel}

\affiliation[a]{Département de Physique Nucléaire et Corpusculaire, Université de Genève, Faculté de Sciences,  1211 Genève, Switzerland}
\affiliation[b]{GSSI - Gran Sasso Science Institute, Via F. Crispi 7, I-67100, L'Aquila, Italy}
\affiliation[c]{INFN - Laboratori Nazionali del Gran Sasso, Via G. Acitelli 22, I-67100, Assergi, L'Aquila, Italy}

\affiliation[d]{Fondazione Bruno Kessler, Via Sommarive, 18, Povo
38123 Trento}
\affiliation[e]{IFJ PAN, PL-31342 Krakow, Poland}

\emailAdd{caterina.trimarelli@gssi.it$^*$, shideh.davarpanah@unige.ch$^*$}

\abstract{
Space missions require lightweight, low-power consuming, radiation-tolerant components. Silicon photomultipliers (SiPMs) are increasingly used for detecting near-UV, optical, and infrared light in space due to their compact design, low cost, low power consumption, robustness, and high photo-detection efficiency, which makes them sensitive to single photons. Although SiPMs outperform traditional photomultiplier tubes in many areas, concerns about their radiation tolerance and noise remain. In this study, we estimate the radiation effects on a satellite in sun-synchronous low Earth orbit (LEO) at 
an altitude of 550~km during the declining phase of solar cycle 25 (2026–2029). We evaluated silicon photomultipliers produced by the Foundation Bruno Kessler (FBK) using front-side illuminated technology with metal trenches (NUV-HD-MT), assessing their response to a 50~MeV proton beam and exposure to a $\beta$-radioactive source (strontium-90). Simulations with SPENVIS and Geant4 were used to validate the experimental results. Based on our findings, we propose a photosensor annealing strategy for space-based instruments.
}

\begin{document}
\maketitle
\flushbottom

%RULESS FOR THE PAPER

% legend with light black box
% for temperature T=
% all the common information title
% micrometer mu
% Font for figures: computer modern (latex font) 

\section{Introduction: silicon photosensors in space applications}
\label{sec:intro}

Silicon photomultipliers (SiPMs) offer high photon detection efficiency (PDE) and low-power consumption, making them ideal candidates for a range of space applications~\cite{Renker:2009zz,Acerbi:2019qgp,Montaruli:2020vjj,Nagai:2019yzb}. One of the key advantages of SiPMs is their ability to operate efficiently in low-light conditions.
%, enabling the detection of faint signals in the vast expanse of space. 
Their compact size and robustness also make them well-suited for integration into lightweight space instruments.

However, a major downside when using SiPMs in space missions is their susceptibility to radiation, which induces damages and degrades their performance over time. In addition, SiPMs are sensitive to temperature variations, requiring meticulous thermal management to ensure reliable operation in a harsh space environment. The work presented here aims to address these limitations in order to enhance the reliability and performance of SiPMs for space applications.

The NUSES~\cite{Trimarelli:2024rdb} project will employ SiPMs for both of its payloads, Terzina~\cite{Burmistrov_2023} and Ziré~\cite{DeMitri:2023ejv}. Ziré, dedicated to monitoring of low-energy ($<250$~MeV) cosmic ray fluxes in the Van Allen belts, gamma-ray bursts in the energy region 1-10~MeV, space
weather and lithosphere-ionosphere-
magnetosphere couplings, will employ SiPMs with LYSO or GAGG crystals in the calorimeter (CALOg), in the plastic scintillator tower (PST) reading scintillating tiles, and in the fiber tracker (FTK).

In this work, we consider the effects of radiation for the Terzina camera~\cite{instruments8010013}. Differently than the Ziré detectors, the camera will be more exposed to the direct effect of radiation as it is not screened by scintillating material.
Here, we present the results of a full characterisation of the sensors of the Fondazione Bruno Kessler (FBK) and an analysis of their response after irradiation to select a suitable sensor for the Terzina telescope~\cite{instruments8010013}. 

Several other space missions are leveraging SiPM technology to advance our understanding of high-energy astrophysical phenomena.
SiPMs (produced by Hamamatsu Photonics) are foreseen to be used for the HERD (High Energy Cosmic Radiation Detection) plastic scintillator detector~\cite{HERD}. HERD is a planned space-based experiment to study dark matter from space through the measurement of energy spectra and anisotropy of high energy electrons, as gamma-ray monitoring between 0.5 GeV and 100 TeV and to measure cosmic ray fluxes to the knee.
A future mission following the paths of Ziré employing scintillating crystals read by SiPMs is Crystal Eye~\cite{Barbato_2019}.

SiPMs play a crucial role in Cherenkov light detection in Terzina. The proposed POEMMA (Probe Of Extreme Multi-Messenger Astrophysics) mission~\cite{Olinto_2021,2023arXiv230914561O} firstly proposed to complement measurements of fluorescence with MAPMTs with Cherenkov detection with the more sensitive SiPMs. Cherenkov light is emitted by ultra-high-energy cosmic rays (UHECRs) and can also be detected from earth-skimming astrophysical neutrinos when they produce extensive air showers (EAS) in the Earth's atmosphere.
This mission is preceded by several balloon flights under the JEM-EUSO (Joint Experiment Missions for Extreme Universe Space Observations) program~\cite{jemeuso,Bacholle_2021}. In particular, EUSO-SPB2 (Extreme Universe Space Observatory on a Super Pressure Balloon 2)~\cite{adams2017white,Bisconti:2023jnl}, launched in May 2023 from NASA's Wanaka base in New Zealand~\cite{spb2-news}, employed SiPMs, produced by Hamamatsu Photonics, for cosmic-ray detection from the stratosphere.

This study for Terzina, where the photosensor plane is directly exposed to cosmic radiation, is imperative. A previous study was performed on S13361-6075NE-04 and S14161-6050HS-04 from Hamamatsu by POLAR-2~\cite{2023ExA....55..343M}, where SiPMs are coupled with scintillators as for other experiments. They amplify weak signals and provide protection against radiation damage, along with the shielding provided by the satellite structure. The POLAR-2 mission plans to use SiPMs for advanced gamma-ray burst (GRB) polarization studies and radiation hardness studies.
Specifically regarding radiation hardness, they have proposed mitigation strategies that align with the work described here~\cite{2023NIMPA104867934D,2023ExA....55..343M}.

The paper is structured as follows: in section~\ref{sec:terzinaplane}, the Terzina telescope and its Focal Plane Assembly (FPA) are described. In section~\ref{sec:charact}, we detail the results of the single SiPM characterisation that will compose the tiles of the FPA. In section~\ref{sec:spenvis}, we describe the calculation of the fluxes of cosmic rays at the altitude of the NUSES orbit with the SPENVIS software~\cite{SPENVIS}. In section~\ref{sec:geant4}, we estimated the ionizing and non-ionizing doses affecting the FPA through a Geant4-based~\cite{GEANT4,ALLISON2016186} simulation of the detector. In sections~\ref{sec:protons} and~\ref{sec:thermic}, a proton irradiation campaign of the SiPMs that will be used in the Terzina telescope is described together with the thermal response of the sensors. Moreover, the effect of the dose released by electrons has been taken into account (section~\ref{sec:electrons}). Finally, a discussion of the results and of a possible mitigation strategy for radiation damage are provided in section~\ref{sec:9}.

\begin{figure}[t!]
    \centering      
    \includegraphics[width=0.8\textwidth]{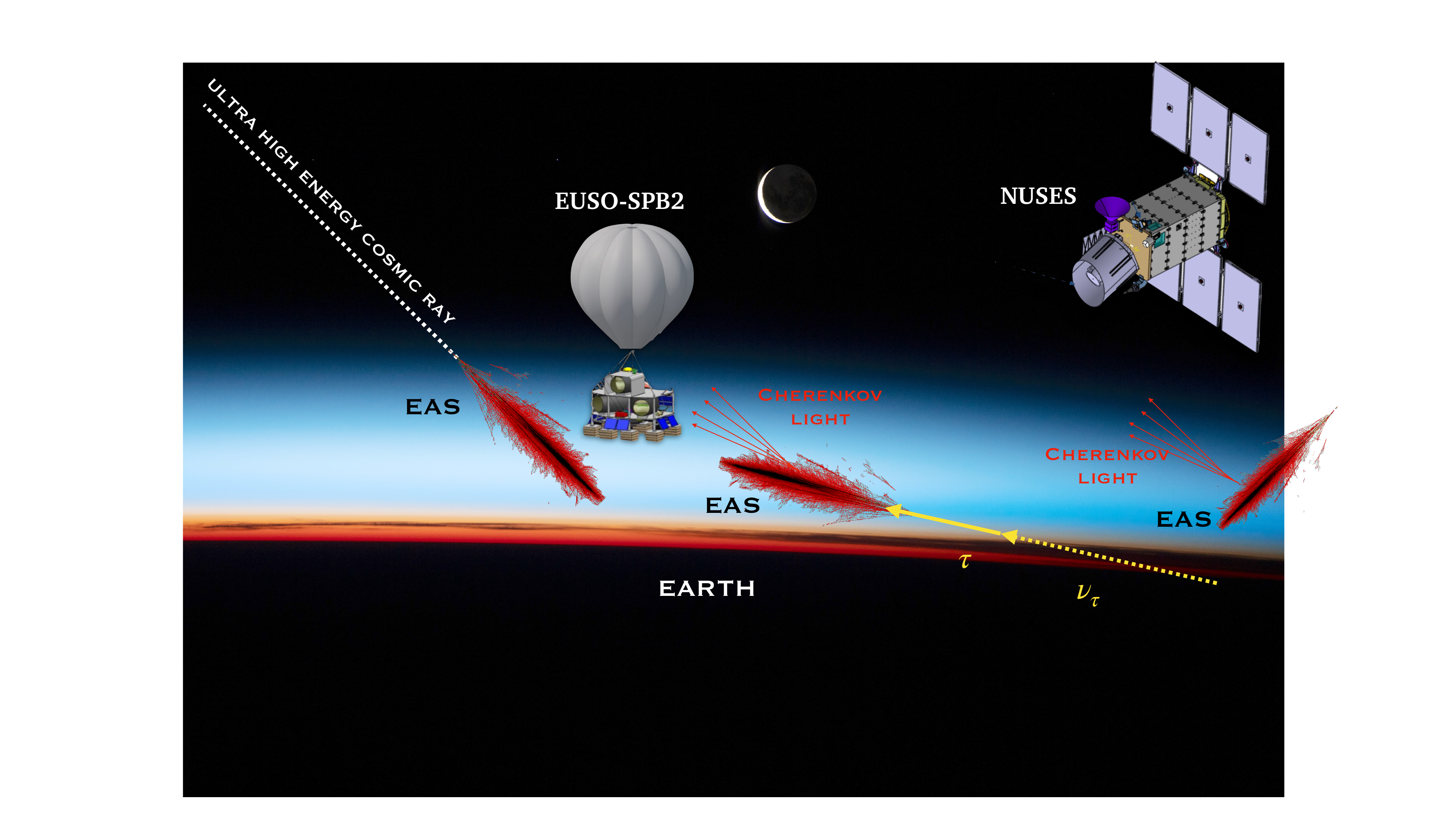}
        \caption{\label{fig:NUSES} A pictorial view of EAS detection by space-based instruments such as Terzina on board NUSES.}
\end{figure}

\section{The Terzina telescope onboard NUSES}
\label{sec:terzinaplane}

The NUSES ballistic mission aims to explore new scientific and technological pathways for future space-based astroparticle physics detectors. The satellite platform, developed by the industrial partner TAS-I (THALES Alenia Space - Italia), will operate in a quasi-polar Low Earth Orbit (LEO) at an altitude of 535~km in the worst case and 550~km at the Beginning of Life (BoL). The satellite will host two instruments: Terzina and Ziré. It will follow a sun-synchronous orbit and operate in a dusk-dawn mode along the day/night boundary so that the Terzina telescope can point to the dark side of the Earth's limb to detect the faint Cherenkov light from ultra-relativistic particle showers in the dark atmosphere. The Terzina telescope aims at the first detection of this Cherenkov light, which can be detected within the field of view (FoV) of its camera located in the FPA. The FPA is designed to detect the Cherenkov light emitted by UHECR-induced EAS with energies above $\sim100$~PeV and with their axis pointing upwards close to the optical axis of the telescope pointing to the limb. 

The Earth's limb is at the angle $\theta_{\mathrm{nadir}} = \arcsin \frac{R_\oplus}{R_\oplus + h}$, where $\theta_{\mathrm{nadir}}=67.04^\circ$ when $h = 550$~km at BoL while $\theta_{\mathrm{nadir}}=67.34^\circ$ at the end of life, when the drag of the residual atmosphere will reduce the orbit altitude to 535~km. In addition, Terzina can detect the upward EAS produced in the atmosphere by the decay of tau and muon leptons, which are generated by the respective neutrinos with energies above $\sim 100$~PeV, passing through the Earth. 
A pictorial view of the detection principle of the Terzina payload on the NUSES satellite is presented in figure~\ref{fig:NUSES}, highlighting the instrument's positions and pointing directions, though not in scale. Terzina is a pathfinder for future larger missions, such as POEMMA or a constellation of satellites as conceived for CHANT~\cite{PhysRevD.95.023004}. 

The Terzina detector consists of a near-UV optical telescope with Schmidt-Cassegrain optics (shown in figure~\ref{fig:optic_scheme}) with effective collective area of 0.09~m$^2$ and equivalent focal length of 925~mm and a FPA in the focal plane. The shadowing due to the support of the system (baffles and vanes) is 10\%. The Optical Telescope Assembly (OTA), which will be realised by the company Officina Stellare, is composed of a hyperbolic primary mirror of 430~mm diameter and radius of curvature (RoC) of 1200~mm, fitting a flat FPA of 131$\times $60~mm$^2$.
The secondary aspherical mirror has a diameter of 194~mm and RoC\,=\,605~mm and it is located at a distance of 430~mm from the primary mirror. The secondary mirror is covered by a correcting lens of thickness 30~mm and RoC\,=\,320~mm to compensate for all aberrations and principally for those induced by the flat FPA (if curved, it would have reduced aberrations but the mechanical design of the FPA would have been more complex).

\begin{figure}[t!]
    \centering
 \includegraphics[width=0.56\textwidth]{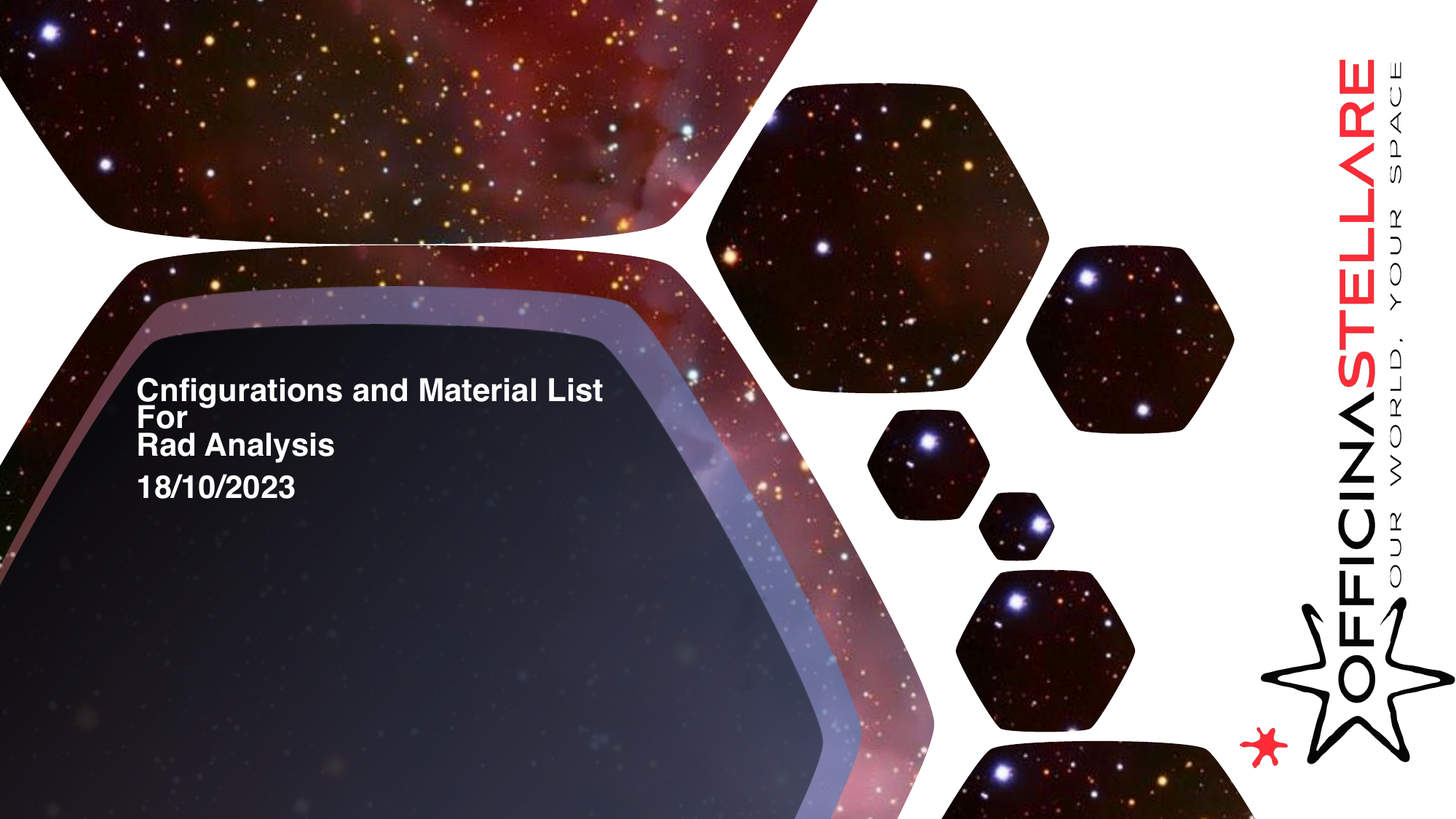}
       \includegraphics[width=0.4\textwidth]{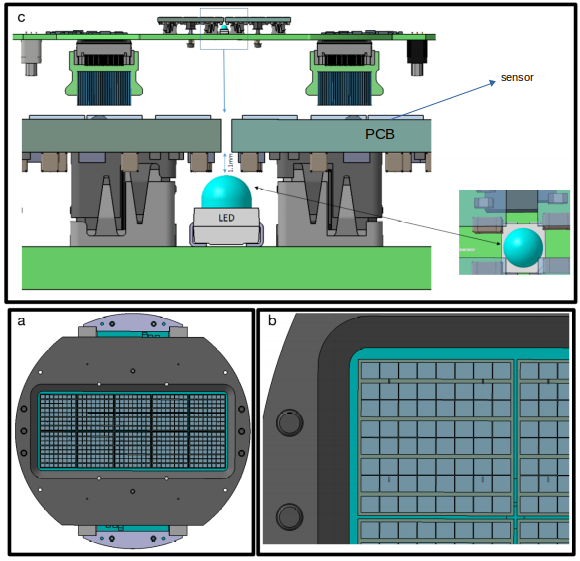}
        \caption{Left panel: cross-sectional view of the preliminary geometry of the telescope.
        Right panel: (a) CAD of the FPA with two rows of 5 tiles occupying an area of $126.5 \times 50.3$~mm$^2$. (b) Detailed view of one tile composed of 64 pixels (c) side view of two tiles and LED position for calibration. \label{fig:FPA_cad}} 
\end{figure}

\begin{figure}[t!]
    \centering

       \includegraphics[width=1.\textwidth]{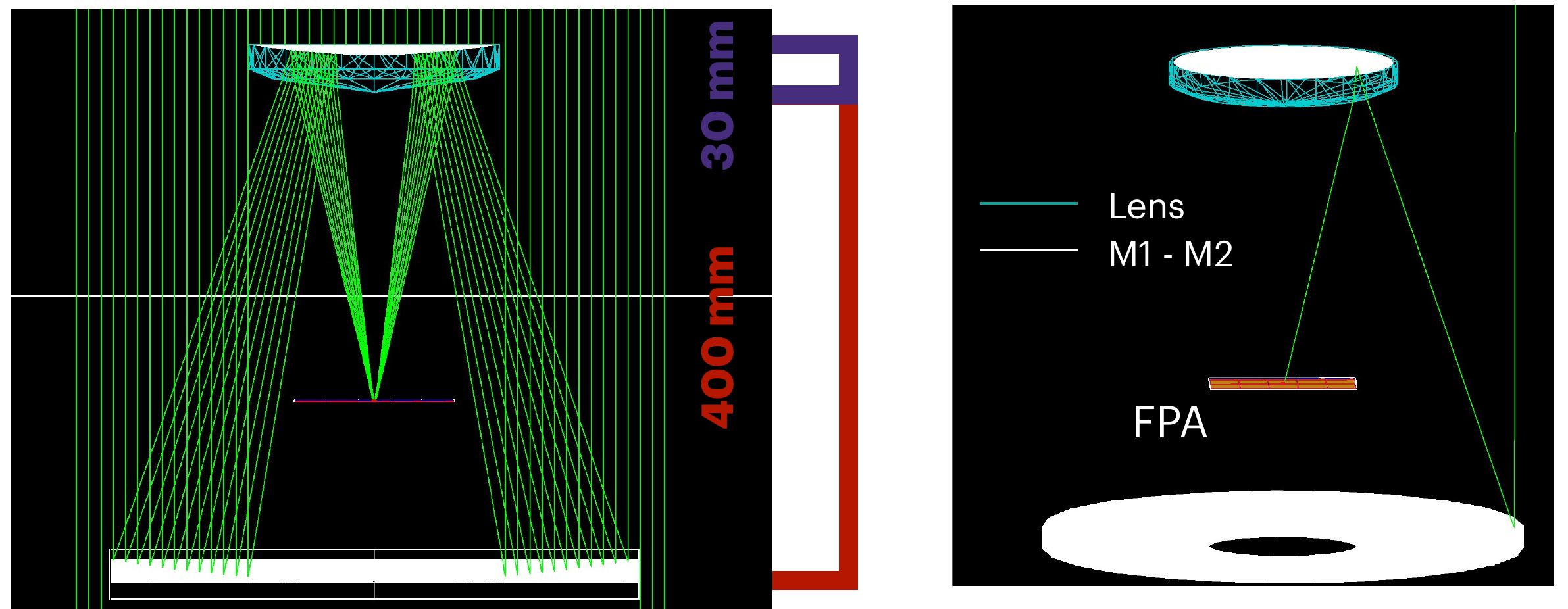}
        \caption{ \label{fig:optic_scheme}Preliminary optical layout of the Terzina telescope. The dual mirror configuration is shown based on the Schmidt-Cassegrain optics. } 
\end{figure}

The FPA is designed to detect photons from both below and above the limb and consists of 10 SiPM arrays, each with $8\times 8$ pixels of $3\times3$~mm$^2$ (640 pixels in total). Each pixel has a sensitive area of $2.73\times2.34$~mm$^2$. As it is known that a refractor and a Cassegrain optical system produce an inverted image when used
without a diagonal mirror~\cite{1944MNRAS.104...48L}, the optical system of Terzina will image the Earth below the limb in the upper row of 5 SiPM arrays of the camera. These will be sensitive to potential neutrino events from below the Earth's limb, though the expected rate is negligible for Terzina. A larger mission will be required. They will be used for characterisation of luminous noise from the Earth. The lower row of 5 SiPM arrays will observe EAS generated by UHECRs from above the limb. The telescope's FoV is $7.2^\circ$ horizontally and $2.9^\circ$ vertically, as each pixel sees $\sim0.18^\circ$.
A computer assisted design (CAD) model of the Terzina telescope is shown in figure~\ref{fig:FPA_cad} (on the left) and details of the CAD of the FPA and the tiles are shown on the right.

\section{The NUV-HD-MT SiPMs for Terzina}
\label{sec:charact}

The requirements for the SiPM technology to be adopted by Terzina are: 
\begin{enumerate}
\item to achieve PDE\,$>50\%$ at 400~nm, namely in the region of the peak of the Cherenkov light spectrum (which for Terzina depends on the altitude of the EAS, but for the bulk of the events at around 30 km is between 350-500~nm);
\item optical cross-talk of less than 10\% at the operation voltage; 
\item dark count rate (DCR) of less than 100 kHz/mm$^2$ at BoL of the mission. 
\item signal duration of less than 40 ns full width half-maximum (FWHM).
\end{enumerate}
In partnership with FBK, we have selected their technology minimizing the optical cross-talk and the DCR, the NUV-HD-MT~\cite{s19020308}. We then investigated which micro-cell size would maximise the PDE while keeping the signal duration short. In fact, the signal duration is proportional to the micro-cell size and its capacitance. A large signal increases the noise captured within the acquisition window required to integrate the full pulse. The NUV-HD-MT technology has been developed by FBK adding metal-filled Deep Trench Isolation (DTI) to strongly suppress optical cross-talk. FBK finds a reduction in optical cross-talk by a factor of 10~\cite{gola2022} with respect to the NUV-HD in absence of the metal-filled DTI. We have measured the NUV-HD-MT SiPM of $3\times3$~mm$^2$ and $1\times1$~mm$^2$ with square micro-cell sizes of  25, 30, 35, 40, 50~$\mu \text{m}$. Another significant consideration is whether to use bare or coated sensors. While bare sensors require meticulous handling, the electrons may radiate Cherenkov light within the covering resin layer~(see section~\ref{sec:electrons}). 

\subsection{Static characterisation}
All the laboratory measurements have been performed at the University of Geneva, according to the setup and analysis methodology described in \cite{Nagai:2019yzb,Nagai_2019}. Figure~\ref{fig:setup_Schematic} shows the schematic layouts of the experimental setups for static (left) and dynamic/optical characterisation (right). 
The two configurations incorporate the following elements:

\begin{figure}[t!]
    \centering
    \includegraphics[height=6.1cm]{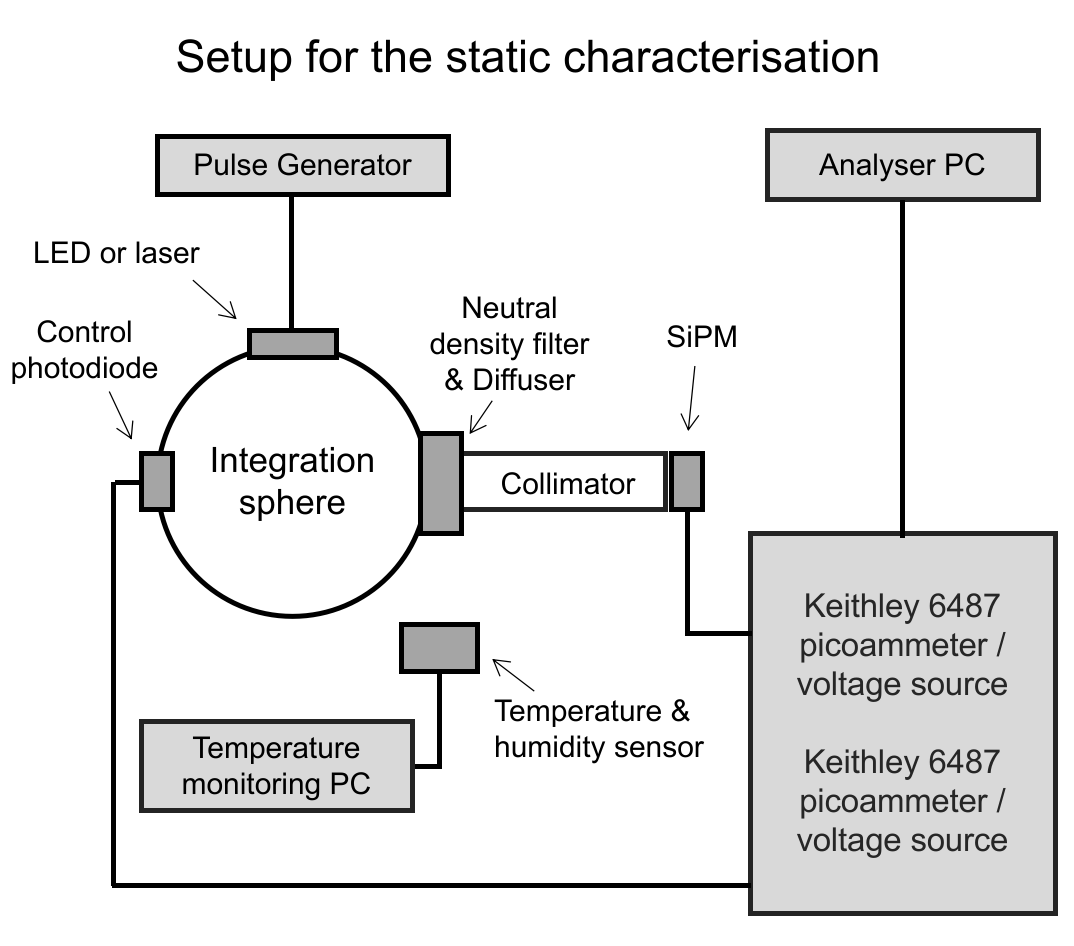}
    \includegraphics[height=6.1cm]{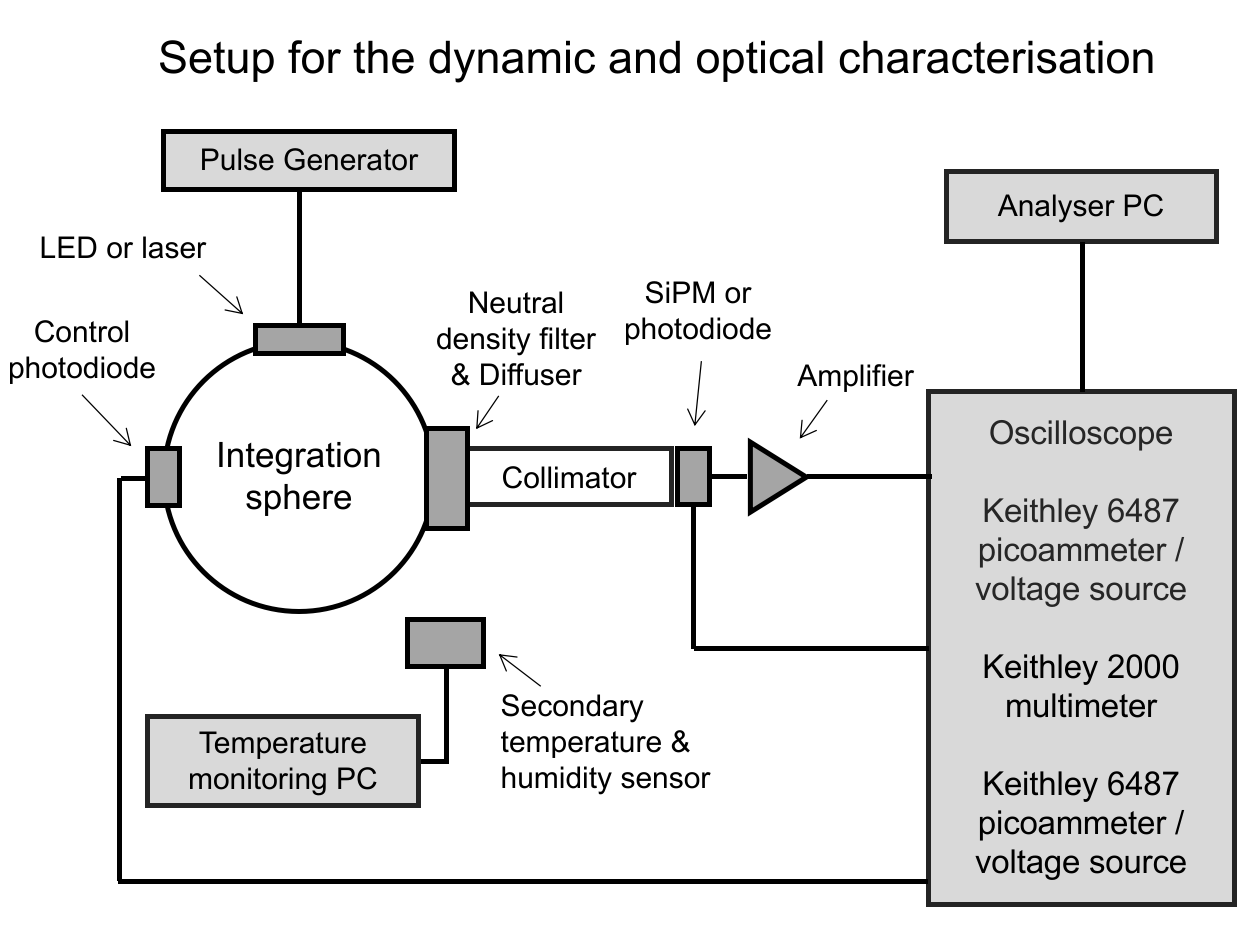}
    \caption{Schematic layouts of the experimental setup for static characterisation (on the left) and the dynamic/optical characterisation (on the right).}
    \label{fig:setup_Schematic}
\end{figure}

\begin{itemize}
    \item a pulsed LED with 450~nm wavelength, powered by a pulse generator operating at the frequency of 10~kHz. This creates alternating intervals: one where the device operates in complete absence of light and another where it is illuminated by the LED light (synchronized with the trigger waveform acquisition), during which photon signals are expected. Hereafter, these time intervals are referred to as \textit{dark} interval and \textit{trigger} window, respectively; 
    \item an integration sphere which eliminates any spatial variations in the incoming light by multiple scattered reflections on its internal surface. This ensures a uniform diffuse light output and preserves the power at each output port;
    \item a  neutral density filter inserted between the integration sphere output port and the SiPM. It is mounted on a motorised wheel for automatic switching between each filter. This allows us to illuminate the SiPMs with different light intensities as required;
    \item a diffuser, inserted immediately after the neutral density filter, to further reduce any residual non-uniformity from the output port of the integration sphere inside the collimator to the sensors;
    \item a control photodiode, mounted on another output port of the integration sphere to monitor the amount of the light within the setup at all times;
    this is connected to a Keithley 6487 picoammeter\footnote{The Keithley 6487 picoammeter is able to perform  measurements of the current with precision of around 6.4~pA for currents lower than 2~nA.}/voltage source~\cite{Ketihley6487};
    \item a temperature and humidity sensor, placed near the SiPM, which constantly monitors the temperature during the measurement;
    \item an analyser PC that monitors and records the output of all instruments.
\end{itemize}

For the static characterisation configuration, the IV curve was measured by directly biasing and reading the SiPM using a Keithley 6487 or Keithley 2400~\cite{Ketihley2400} picoammeter/voltage source. In the dynamic characterisation configuration, the bias voltage was supplied by a Keithley 6487 picoammeter/voltage source reading before the signal with a 2~GHz bandwidth oscilloscope.
A custom-made amplifier with a bandwidth of approximately $\sim600$~MHz and a gain of $\sim40$~dB was used to amplify, pulse shape and filter the SiPM signal, and effectively suppress the slow component of the SiPM response effectively. The Keithley 2000 multimeter~\cite{Ketihley2000}, shown in this configuration, has been used for reading a temperature sensor mounted near the control photodiode. 
The average temperature during the measurements was $T=22$~$ ^{\circ}$C (hereinafter referred to as room temperature). 

The static characterisation has been done: the forward IV measurements have been conducted to estimate the quenching resistance, while the reverse IV measurements have been used to determine the breakdown voltage.

\subsubsection{Forward IV characterisation}

\begin{figure}[t!]
\centering
\quad 
\includegraphics[width=0.42\textwidth]{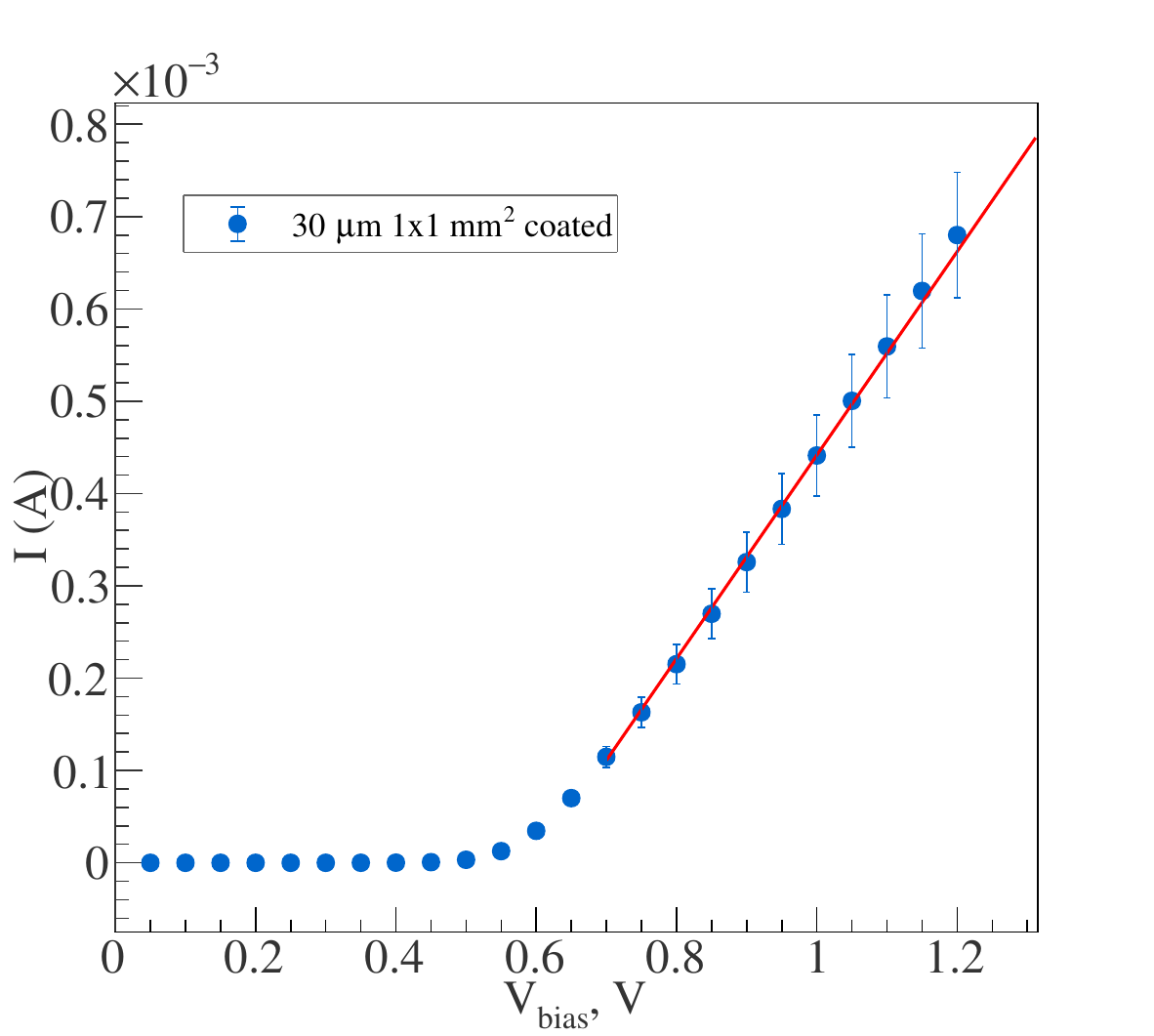}
\quad
\includegraphics[width=0.52\textwidth]{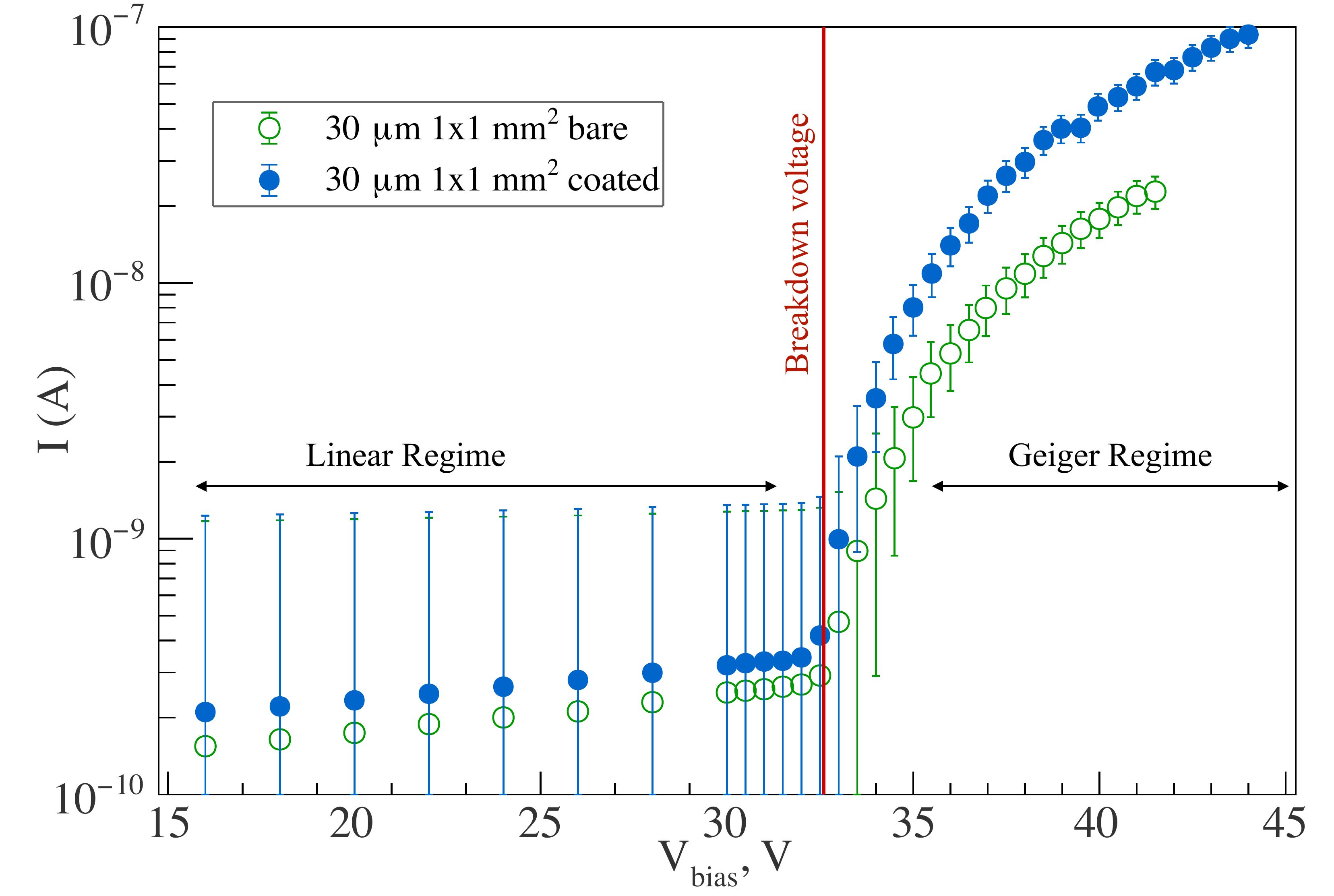}
\caption{The forward (left) and reverse (right) IV characteristic data points with scale errors of the instruments for the FBK SiPM with 30 $\mu \mathrm{m}$ micro-cell size. The errors are due to the instrument scale precision.\label{fig:IV_curve}}
\end{figure}

The forward IV characteristic curve of the FBK SiPMs, shown in figure~\ref{fig:IV_curve} (left panel), exhibits a very small increase in current when the polarization voltage, $V_{\mathrm{bias}}$, is below the threshold value, followed by a rapid, linear increase in current with $V_{\mathrm{bias}}$ above this threshold. This behaviour can be described by the Shockley law, which governs the forward current $I_d$, flowing through a $p$-$n$ junction diode. A SiPM is an array of $\mu$-cells ($N_{\mu \mathrm{cell}}$) connected in parallel, each referred to as a SPAD (single photon avalanche diode). Each $\mu$-cell can be modelled as a diode in series with a quenching resistor, $R_q$. For the entire device, the voltage can be expressed as:
\begin{equation}
V_{\mathrm{bias}}=\eta V_{\mathrm{T} }\left [\ln \left(\frac{I}{I_s}+1\right)\right] +I\frac{(R_s+R_q)}{N_{\mu \mathrm{cell}}} \, .
\end{equation}
Here, $\eta$ is the ideality factor (ranging from $\eta = 1$ for pure diffusion current and $\eta = 2$ for pure recombination current), $I_s$ is the total reverse bias saturation current, $V_{\mathrm{T}}$ is the thermal voltage, typically $R_s=100~\Omega$~\cite{Nagai_2019} and $I$ is the total forward current flowing through the SiPM. When the current is high ($I/N_{\mu \mathrm{cell}}>5~\mu$A), the last term becomes dominant. In this regime, the quenching resistance can be determined from a linear fit of the forward IV characteristic curve, as shown in figure~\ref{fig:IV_curve} (red curve). The measured values of the quenching resistance are shown in table~\ref{tab:forwad_resistor} ($\sim$1000~k$\Omega$) for both bare and coated sensors of the same size but with varying micro-cell dimensions.

\begin{figure}[t!]
\centering
\includegraphics[width=0.94\textwidth]{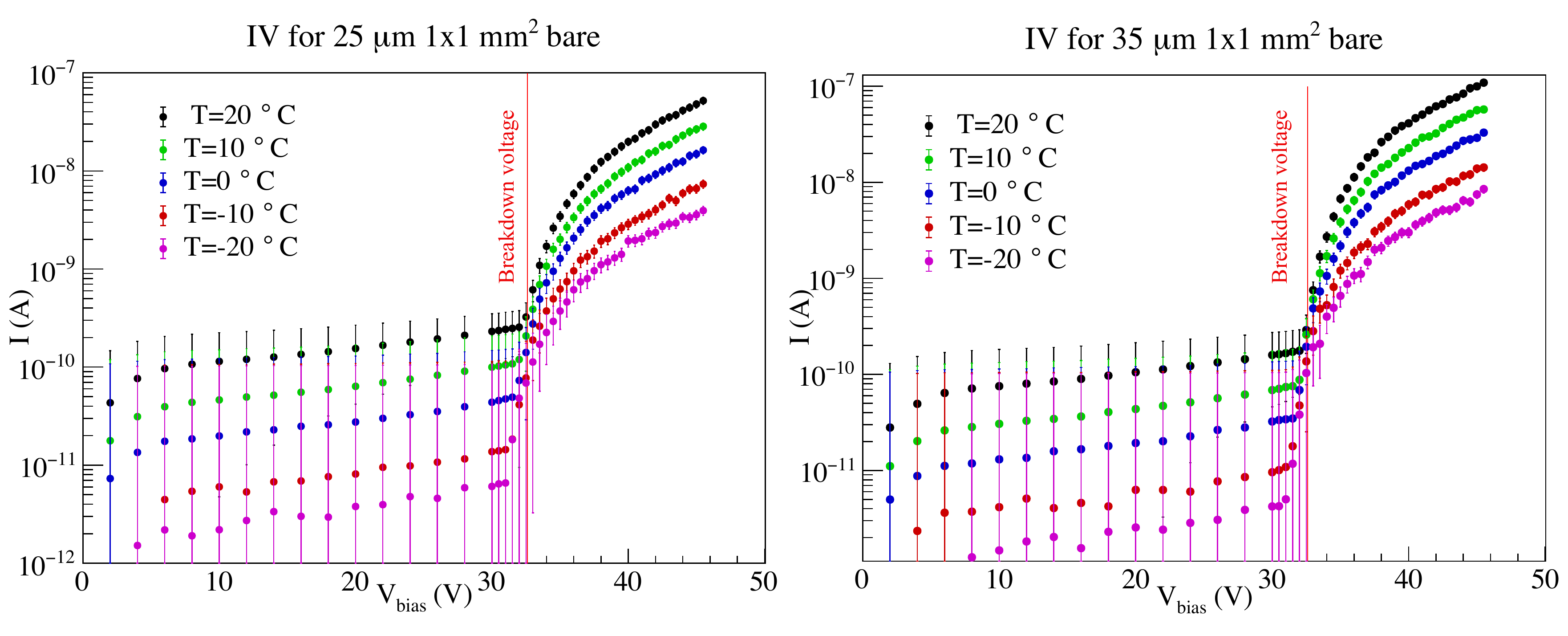}
\caption{The reverse IV characteristic data points with scale errors of the instruments for the bare FBK SiPMs with 25 $\mu \mathrm{m}$ micro-cell size (left) and 35 $\mu \mathrm{m}$ micro-cell size (right) for different temperatures. The errors are due to the instrument scale precision.\label{fig:temp_device}}
\end{figure}
\begin{table}[t]
\centering
\begin{tabular}{|c|c|c|c|c|}
\hline
sensor area (mm$^2$) & cell size ($\mu $m) & $R_q$ (k$\Omega$) & $V_{\mathrm{BD}}$ (V) & Coating\\
\hline
 $1\times1$ & 25   & $950.0 \pm 2.0 $ & $32.60\pm 0.10$ & no\\
  $1\times1$ & 35   & $964.0 \pm 3.0 $ & $32.58\pm 0.10$ &no\\
\hline
  $1\times1$ & 25   & $990.0 \pm 2.0$ & $32.60\pm 0.12$& yes\\
  $1\times1$ & 30   & $1012.0 \pm 2.0$ & $32.61\pm 0.15$ & yes\\
  
\hline
\end{tabular}

\caption{\label{tab:forwad_resistor} Quenching resistance and breakdown voltage estimated at 22$^{\circ}$C by fitting the forward and reverse regions respectively of four coated and bare sensors of $1 \times 1$~mm$^2$ and different micro-cell sizes.}
\end{table}

\subsubsection{Reverse IV characterisation and IV model for post-breakdown} %model 1

The reverse IV characteristic curve of the SiPM sensors is shown in figure~\ref{fig:IV_curve} (right panel). The breakdown voltage, $V_{\mathrm{BD}}$, has been calculated by using the second log-derivative method~\cite{Nagai_2019}. The results, presented in table~\ref{tab:forwad_resistor}, indicate that $V_{\mathrm{BD}}$ is approximately 32.6~V at room temperature for all cases.

\begin{figure}[t!]
\centering
\includegraphics[width=1.\textwidth]{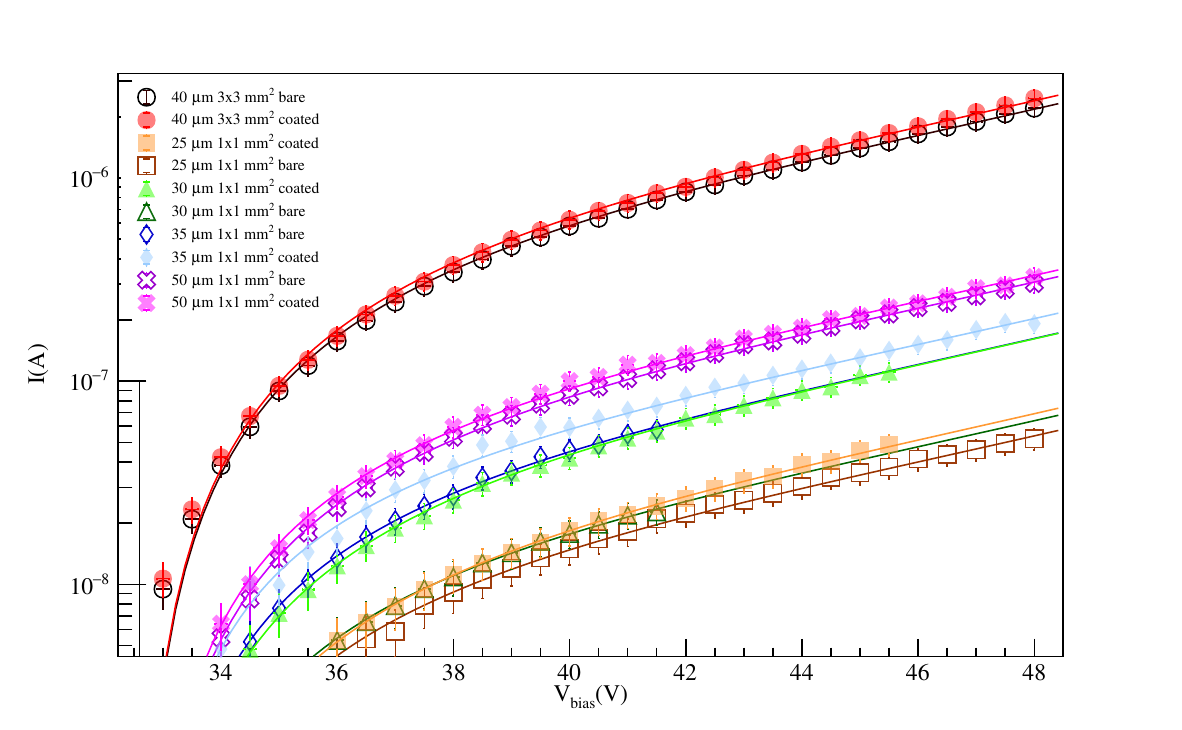}
\caption{Reverse IV characteristics (data symbols) and model (lines) for various SiPM samples at room temperature.\label{fig:IV_postbreakdown_fit}}
\end{figure}

In the reverse IV curves, two distinct regimes can be observed: the pre-breakdown ``linear" regime and the post-breakdown Geiger regime, as illustrated in figure~\ref{fig:IV_curve}. This work employs the so-called ``IV model"~\cite{Dinu:2016hog,Nagai_2019}, which describes the behaviour of SiPMs over their complete operating range, including the Geiger regime. The IV curves for all temperatures and devices exhibit similar behaviour, as it is shown in figure~\ref{fig:temp_device}, indicating multiple regimes of bias voltage. Taking all contributions into account, the post-breakdown current, $I_{\mathrm{post-BD}}$, can be expressed as\footnote{The formula does not include the pre-breakdown term.}:
\begin{equation}
%\begin{split}
I_{\mathrm{post-BD}}(V_{\mathrm{bias}})= \frac{d N_{\mathrm{carriers}}}{dt}\cdot \frac{V_{\mathrm{cr}}-V_{\mathrm{BD}}}{V_{\mathrm{cr}}-V_{\mathrm{bias}}}\times (1 - e^{[p\cdot(V_{\mathrm{bias}}-V_{\mathrm{BD}})]})\cdot C_{\mu \mathrm{cell}} \times (V_{\mathrm{bias}}-V_{\mathrm{BD}})
%\end{split}
\end{equation}

where $\frac{dN_{\mathrm{\mathrm{carriers}}}}{dt}$ corresponds to the contribution of the carrier-induced current, $V_{\mathrm{cr}}$ denotes the voltage where the after-pulses probability is one, $V_{\mathrm{BD}}$ is the breakdown voltage, $C_{\mu \mathrm{cell}}$ is the capacitance and $p$ is related to the Geiger probability. They are free parameters of the model. The $V_{\mathrm{BD}}$ has been estimated with the second log-derivative method and it is a fixed value in the fit. This formula has been used to fit the experimental IV characteristics as shown in figure~\ref{fig:IV_postbreakdown_fit}, and the corresponding fit values are summarized in table~\ref{tab:fit_postbreakdown}. This model will be used in the next sections to predict IV behaviour without repeating measurements.

\begin{table}[t!]
\centering
\resizebox{\textwidth}{!}{
\begin{tabular}{|c|c|c|c|c|c|c|}
\hline
sensor size & cell size & $dN_{\mathrm{carriers}}/dt$ & $V_{\mathrm{cr}}\,\mathrm{(V)}$ &  $V_{\mathrm{BD}}\,\mathrm{(V)}$ & $C_{\mu \mathrm{cell}}\,\mathrm{(nF)}$ & $p\, \mathrm{(V)^{-1}}$ \\
\hline
 $1\times1$~mm$^2$ & 25~$\mu $m \small{coated} & $0.47 \pm 0.30$ & $60.0$ & $32.6$ & $4.21\pm 0.86$ & $(-3.34\pm 0.32)\cdot10^{-1}$\\
 $1\times1$~mm$^2$ &  30~$\mu $m \small{coated} & $1.00 \pm 0.37$ & $60.0$ & $32.6$ & $4.60 \pm0.89$ & $(-3.56\pm 0.66)\cdot10^{-1}$\\
 $1\times1$~mm$^2$ &  35~$\mu $m \small{coated} & $1.26 \pm 0.39$ & $60.0$ & $32.6$ & $4.59\pm0.86$ & $(-5.95\pm 0.99)\cdot10^{-1}$\\
 $1\times1$~mm$^2$ &  50~$\mu $m \small{coated} & $1.93 \pm 0.55$ & $60.0$ & $32.6$ & $4.88\pm 0.85$ & $(-4.27\pm 0.54)\cdot10^{-1}$\\
 \hline
 $1\times1$~mm$^2$ &  25~$\mu $m bare & $0.54 \pm 0.21$ & $60.0$ & $32.6$ & $2.85\pm 0.95$ & $(-4.07\pm 0.98)\cdot10^{-1}$ \\
 $1\times1$~mm$^2$ &  30~$\mu $m bare & $0.59 \pm 0.40$ & $60.0$ & $32.6$ & $3.08\pm0.98$ & $(-4.33\pm0.99)\cdot10^{-1}$\\
 $1\times1$~mm$^2$ & 35~$\mu $m bare & $1.22 \pm 0.61$ & $60.0$ & $32.6$ & $3.77\pm 0.98$ & $(-4.34\pm0.99)\cdot10^{-1}$\\
 $1\times1$~mm$^2$ & 50~$\mu $m bare & $2.03 \pm 0.58$ & $60.0$ & $32.6$ & $4.32\pm 0.97$ & $(-3.81\pm0.47)\cdot10^{-1}$\\
\hline
$3\times3$~mm$^2$ &  40~$\mu $m bare & $16.9 \pm 4.3$ & $60.0$ & $32.6$  & $ 3.69\pm 0.91$ & $(-3.42\pm0.33)\cdot10^{-1}$\\
$3\times3$~mm$^2$ &  40~$\mu $m \small{coated} & $17.8 \pm 4.5$ & $60.0$ & $32.6$ & $3.86\pm 0.94$ & $(-3.25\pm0.32)\cdot10^{-1}$\\

\hline
\end{tabular}
}
\caption{\label{tab:fit_postbreakdown} Fit results for the post-breakdown IV model measured at 22$^{\circ}$C. For all the fits, the reduced $\chi^2$ (the $\chi^2$ value divided by the number of degrees of freedom) has been estimated by MINUIT~\cite{James:1994vla} .}
\end{table}

\subsection{Dynamic characterisation}
\label{dynamic_characterisation}

To understand and parametrise the dynamic behaviour of the SiPMs for our application, we analysed two categories of noise:
\begin{enumerate}

\item {\bf Primary uncorrelated noise (dark count rate, DCR)}\\
This noise is independent of the light conditions. At room temperature, DCR is mainly caused by the thermal generation of carriers. Under the operation voltage ($V_{\text{bias}}$) and influenced by the trap-assisted tunneling effect, these carriers transition from the valence band to the conduction band, resulting in spurious signals. The increase in the DCR as a function of $V_{\text{bias}}$, due to the increased electric field, can be modelled by the empirical formula:

\begin{equation}
\text{DCR} = N_{\text{carriers}} \times P_{\text{G}}^{\text{DCR}} \times e^{b \times V_{\text{bias}}},
\end{equation}

where $N_{\text{carriers}}$ is the rate of generated carriers at the given temperature, $P_{\text{G}}^{\text{DCR}}$ represents the probability that a generated carrier reaches the high-field region and triggers an avalanche (Geiger probability for dark pulses), and $b$ is a free parameter depending on the impact of $V_{\text{bias}}$.

\item {\bf Secondary correlated noise}\\
This noise arises from secondary avalanches occurring within the same or neighbouring micro-cells and can be categorised into:
\begin{itemize}
    \item {\bf Prompt optical cross-talk (OCT):} triggered by photons emitted during the primary avalanche multiplication process due to hot carrier luminescence.
    \item {\bf Delayed optical cross-talk:} caused by diffuse charge carriers created by photon absorption in non-depleted regions, which then drift through the depleted region.
    \item {\bf After-pulsing (AP):} triggered by carriers trapped in the micro-cell's junction depletion layer during the primary avalanche and released after some delay.
\end{itemize}

In waveforms, prompt and delayed cross-talk appear as an increase in signal (doubling or more) in neighbouring micro-cells. In contrast, after-pulsing manifests within the same micro-cell and is strongly dependent on the recovery time of the SiPM micro-cells.
\end{enumerate}

\begin{figure}[t!]
\centering
\includegraphics[width=0.47\textwidth]{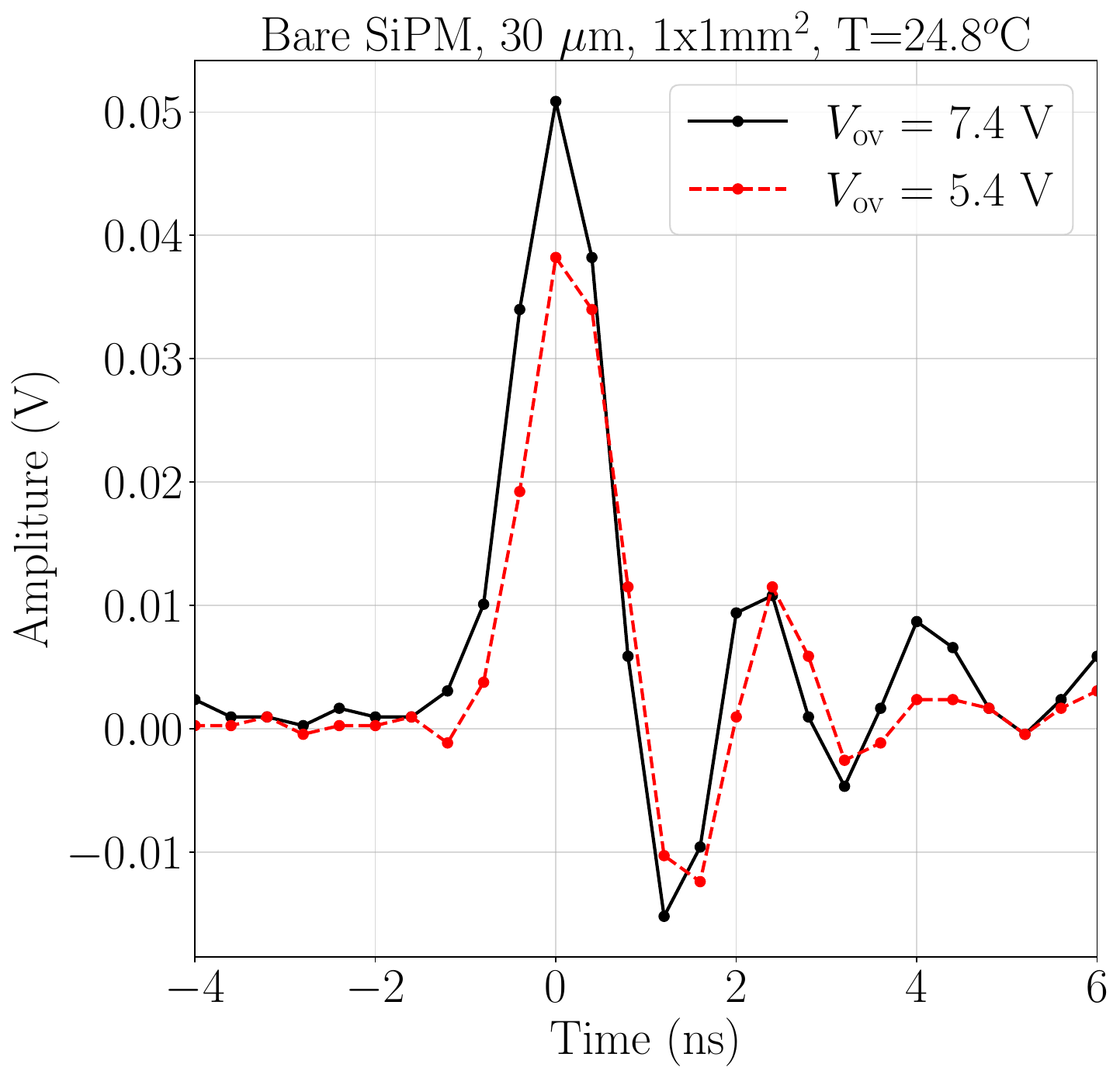}
\caption{\label{fig:sigshape} Single photoelectron response after the preamplifier for different over-voltages for a bare sensor of 30 $\mu$m micro-cell size.}
\end{figure}

To perform the dynamic characterisation of SiPM and measure the uncorrelated noise, we used the setup shown in the right panel of figure~\ref{fig:setup_Schematic}. A full over-voltage scan was conducted by ranging the voltage between 1~V and 18~V, and by acquiring 10'000 waveforms on the oscilloscope for each given over-voltage. To compute the DCR and OCT, the analysis specifically focused on the signals received within \textit{dark} interval of the waveform (in the absence of light). 

Figure~\ref{fig:sigshape} shows a typical single photoelectron signal at the over-voltage of 7.4~V and 5.4~V with full width half-maximum (FWHM) of about 1~ns and amplitudes of 50~mV and 40~mV, respectively. To determine the OCT, we considered the rate of accidental pile-up of 2 or more pulses within the same \textit{dark} interval. A custom-made amplifier with a bandwidth of $\sim600$~MHz and a gain of $\sim40$~dB has been used for dynamic and optical characterisations. It performs pulse shaping and filtering, effectively cancelling the slow component of the SiPM response.

\begin{figure}[t!]
\centering
\subfloat[]{\includegraphics[width=0.47\textwidth]{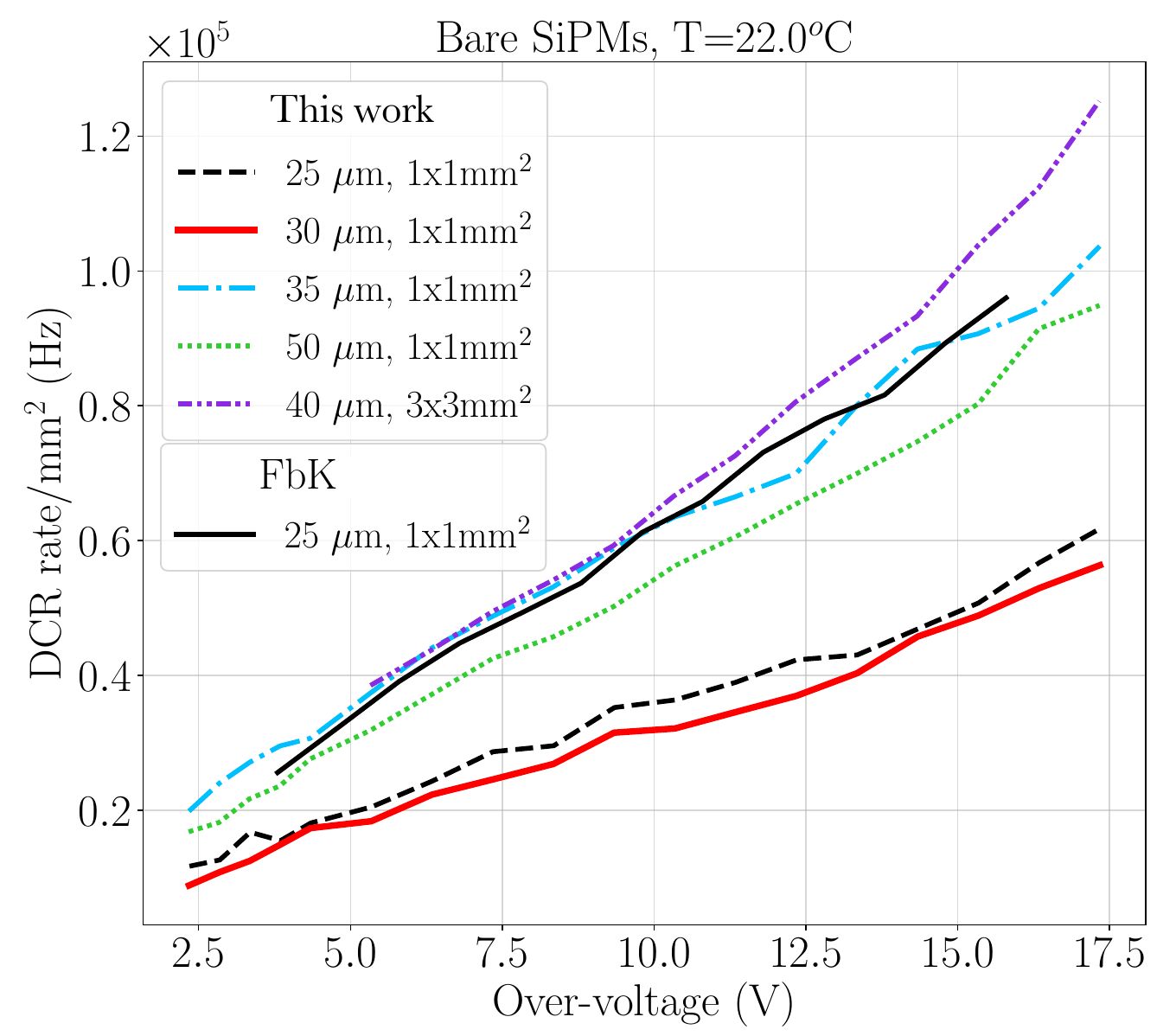}}\qquad
\subfloat[]{\includegraphics[width=0.47\textwidth]{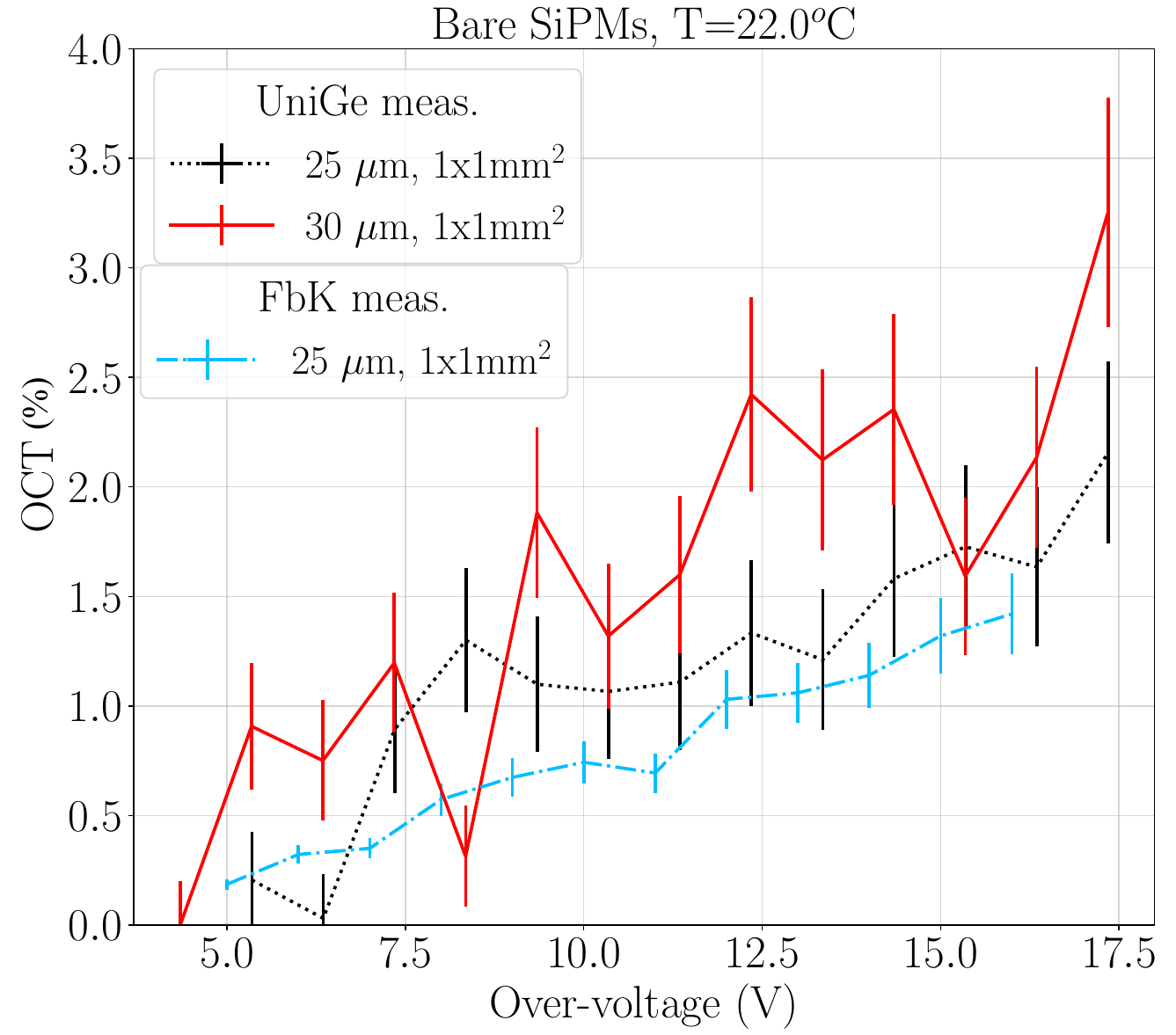}}
\caption{\label{fig:oct} a) Characterisation measurements of DCR per mm$^2$ as a function of the over-voltage at room temperature for sensors without coating with different micro-cell sizes compared to FBK's measurements. b) Our and FBK measurements of OCT as a function of over-voltage for sensors of 25 and 30~$\mu$m micro-cell sizes without coating at room temperature.}

\end{figure}

The results for the DCR and OCT are shown in figure~\ref{fig:oct} in the left and right panels, respectively. At 10 V over-voltage, the DCR per mm$^2$ is less than 60~kHz for all the $\mu$-cell sizes and the OCT is less than $2\%$ for 25 and 30~$\mu$m. Because the OCT is quite small and statistics is limited, the errors are large for our measurements. Our measurements are compared to FBK ones which have a better precision. In addition, our laboratory is close to the RTS radio-television, and affected by a background at their frequencies (DAB$+$ at 223.936~MHz). Hence, we had to screen it by adopting a Faraday cage, but some noise can still be picked up by the cables. Overall, the measurements are in agreement.
 
\subsection{Optical characterisation}
\label{optical_characterisation}

The photon detection efficiency (PDE) is a crucial property of SiPMs, quantifying how effectively the device detects incoming photons as a function of their wavelength ($\lambda$). The PDE can be described as:

\begin{equation}
\mathrm{PDE} = \mathrm{ QE(\lambda) \times \epsilon \times P_{G}(V_{\mathrm{OV}}, \, \lambda)}
\end{equation}

Where $\mathrm{QE(\lambda)}$ is the quantum efficiency, i.e. the probability that a photon generates an electron-hole pair, $\epsilon$ is the fill factor, representing the percentage of the sensor area sensitive to light\footnote{The fill factor is influenced by the $\mu$-cell sizes and their effective area at the pixel level. Specifically, the fill factor depends on how much of the total pixel area is occupied by the $\mu$-cells compared to non-active regions (such as gaps or interconnects) within the pixel.} and $P_{\text{G}}(V_{\mathrm{OV}}, \lambda)$ is the Geiger probability. The PDE strongly depends on the wavelength of the incident light, the over-voltage, and consequently, the breakdown voltage. In addition, PDE is temperature-dependent, as carrier mobility increases with temperature, potentially leading to an increase in $P_{\mathrm{G}}$ for a fixed over-voltage. However, at around room temperature, this effect is likely negligible.
Nonetheless, if the over-voltage is not adjusted to compensate for temperature variations, PDE measurements will fluctuate along with the breakdown voltage.

\begin{figure}[t!]
\centering
\subfloat[\label{fig:pde_all_a}]{\includegraphics[width=0.42\textwidth]
{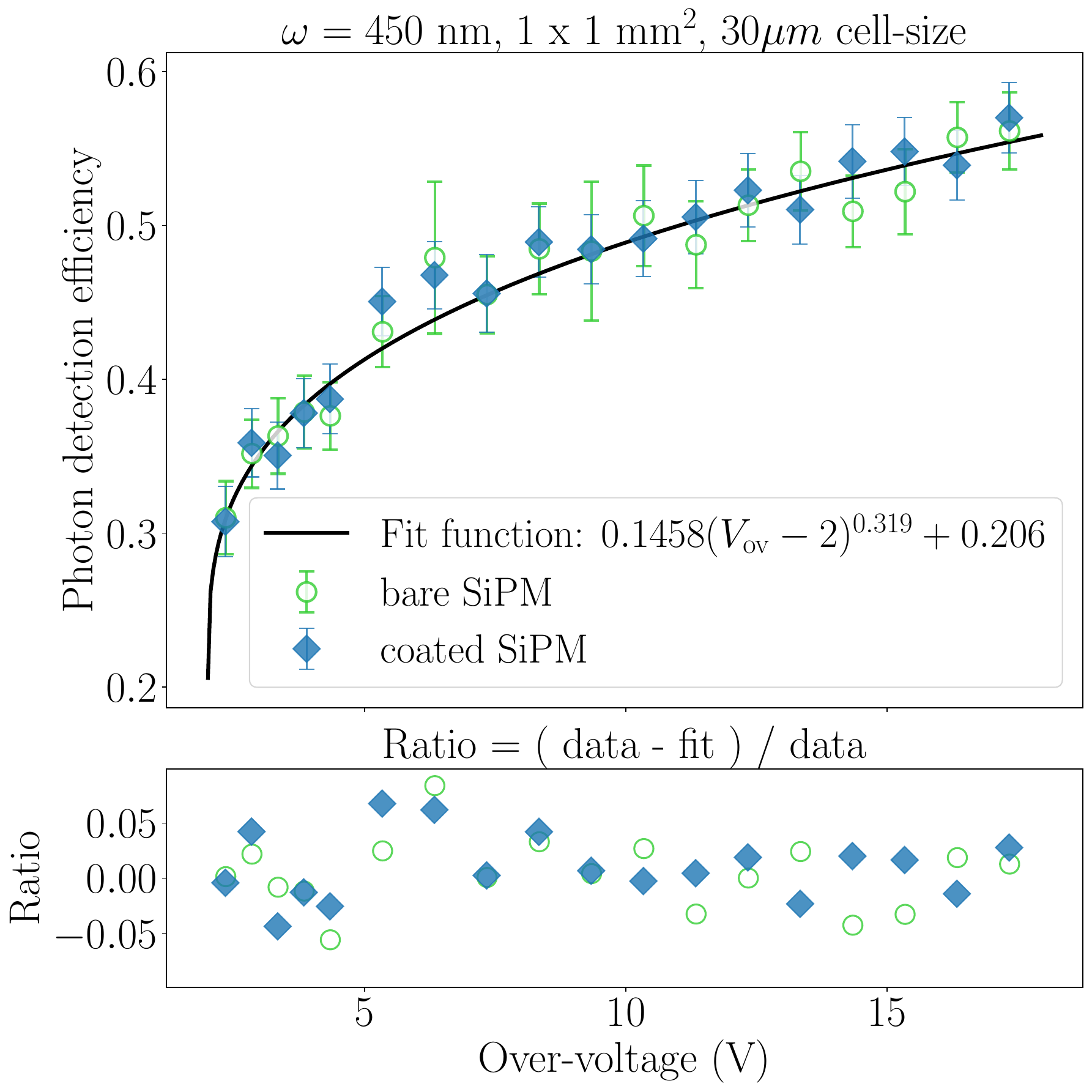}}\qquad
\subfloat[\label{fig:pde_all_b}]{\includegraphics[width=0.47\textwidth]{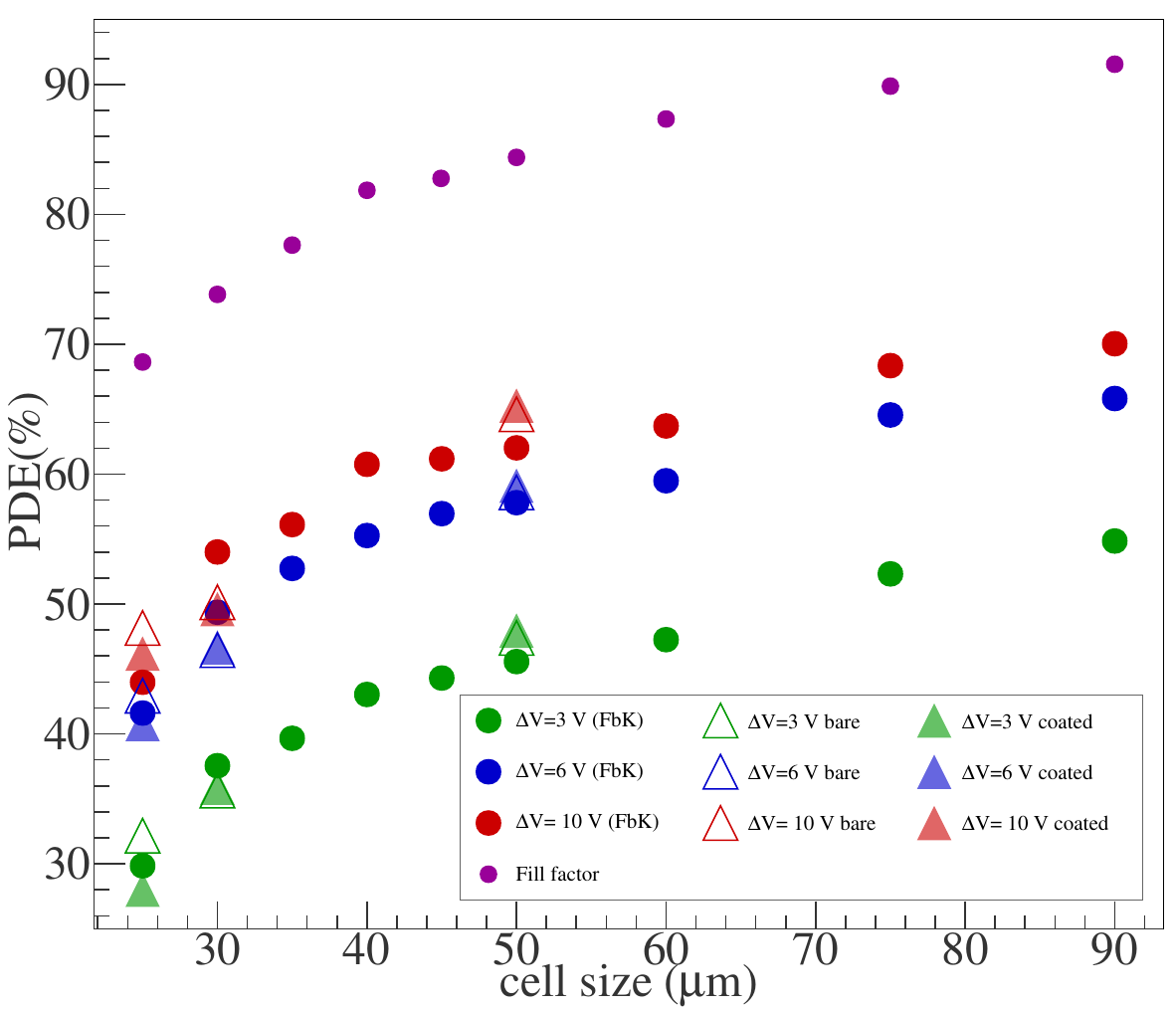}}
\caption{\label{fig:pde_all}a) PDE as a function of over-voltage for 30~$\mu$m cell size without coating. Also, the ratio between the data and the fit is shown in the lower panel. Uncertainties are combined in quadrature, including both statistical and systematic contributions. b) PDE vs the cell size for different over-voltages and for bare and coated sensors. In addition to the measurements made in Geneva at 450~nm, the FBK measurements are shown for 420~nm (full circles).}
\end{figure}

Using the same setup as in the dynamic characterisation (right panel of figure~\ref{fig:setup_Schematic}), we performed a calibration step by replacing the SiPM with a calibrated control photodiode to determine the absolute amount of incoming light. After reinstalling the FBK SiPM, we repeated the scan while illuminating it with a low-intensity pulsed LED. Following the analysis methodology of \cite{Nagai_2019}, we focused on the \textit{trigger} window (the expected arrival time of photons from the LED). The PDE was calculated using the Poisson method, based on the average number of detected photons and corrected for uncorrelated SiPM noise. Figure~\ref{fig:pde_all_b} shows the PDE as a function of the cell size for different over-voltages and sensor coatings. By looking at this plot and considering the results of the OCT and DCR, the 30~$\mu$m cell size bare sensor appears to be the best solution for a LEO satellite. At 10~V of over-voltage, in fact, the PDE for 30~$\mu$m cell size is already more than 50$\%$ (see figure~\ref{fig:pde_all_a}).

\begin{figure}[t!]
\centering
\includegraphics[width=.7\textwidth]{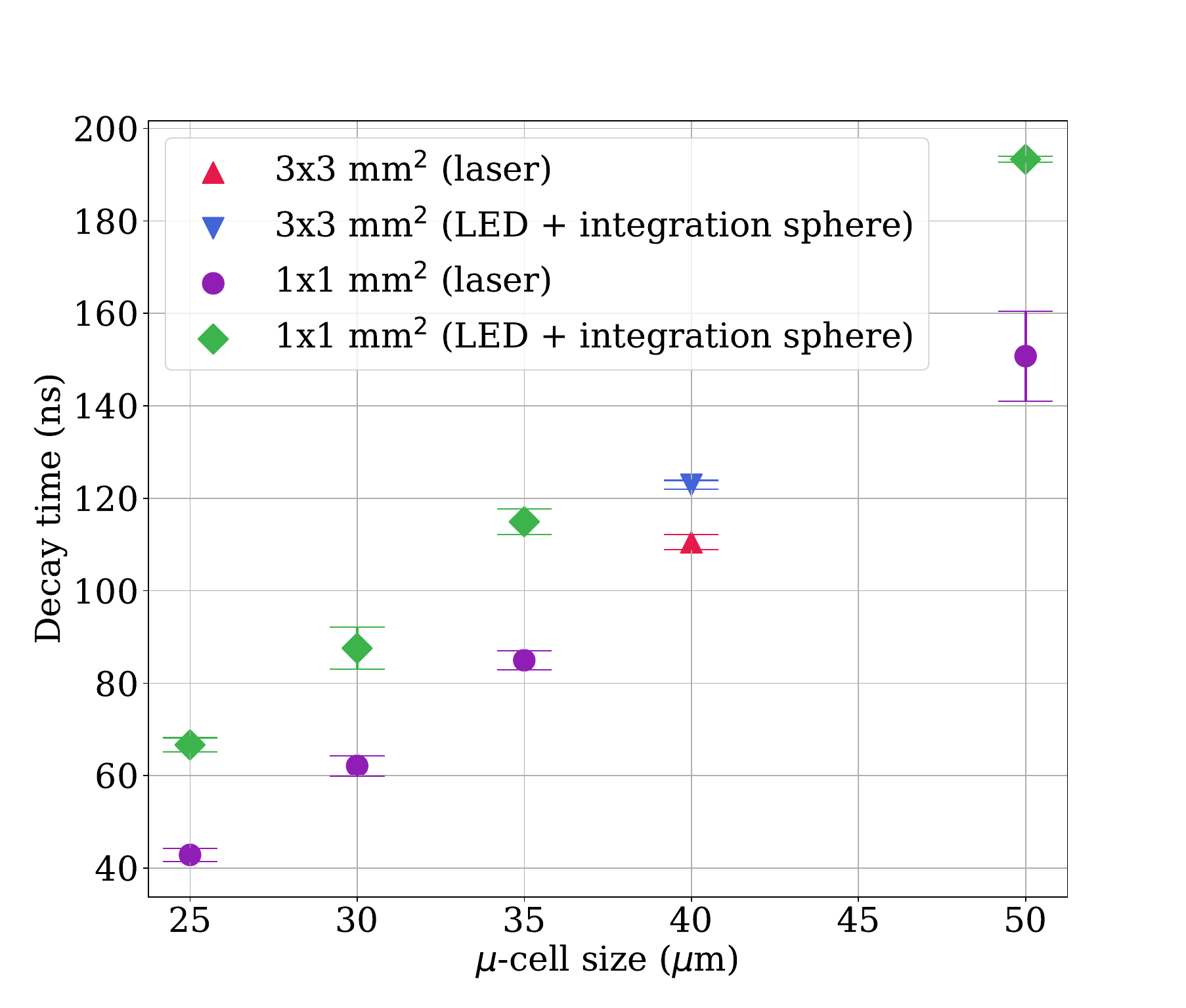}
\caption{SiPM signal decay time as a function of the $\mu$-cell size measured with a LED (450~nm) and a short pulsed laser (370~nm), respectively with and without an integration sphere. 
\label{fig:diff_cell}}
\end{figure}

Weighing the DCR and PDE, as described above, against the pulse shape as outlined in~\cite{instruments8010013}, it has been decided to use SiPM tiles with a 30~$\mu$m bare $\mu$-cell size. With this configuration, we achieve optimal performance in terms of DCR and PDE while maintaining a short pulse decay time.
The decay time and the PDE are anti-correlated: smaller $\mu$-cell sizes result in shorter signal durations but also in lower PDE due to the reduced fill factor. Hence, the final choice of the SiPM $\mu$-cell of 30~$\mu$m represents a trade-off between these parameters. Additionally it complies with the constraints on the maximum power consumption of the NUSES satellite. As a matter of fact, the larger the DCR, the larger the power consumption of the SiPM power supplies. If the data acquisition thresholds are not adjusted accordingly, the rate of noise induced triggers would also require more processing activity, further increasing power consumption.
We performed a set of measurements with three different configurations: 1) with a pulsed LED (450~nm) and an integration sphere; 2) with a short-duration ($\sim$25~ps) light-pulse laser with a 370~nm wavelength laser with and 3) without the integration sphere. Both the LED and the laser flash the full sensor to obtain a high amplitude signal to be detectable by a 2~GHz bandwidth oscilloscope. Figure~\ref{fig:diff_cell} summarises the obtained results. As expected, the shortest decay time was measured with the laser only, as the integration sphere induces an additional time spread.
The SiPM signal tail has been fitted with a negative exponential function and the decay time is $\tau$ in the figure. This time is a function of the $\mu$-cell capacitance, hence its size. We did not observe any significant change in the signal shape with the variation in the bias voltage.

\section{Evaluation of radiation fluxes for a LEO space mission} 
\label{sec:spenvis}

\begin{figure}[t!]
\centering
\quad 
\includegraphics[width=1.\textwidth]{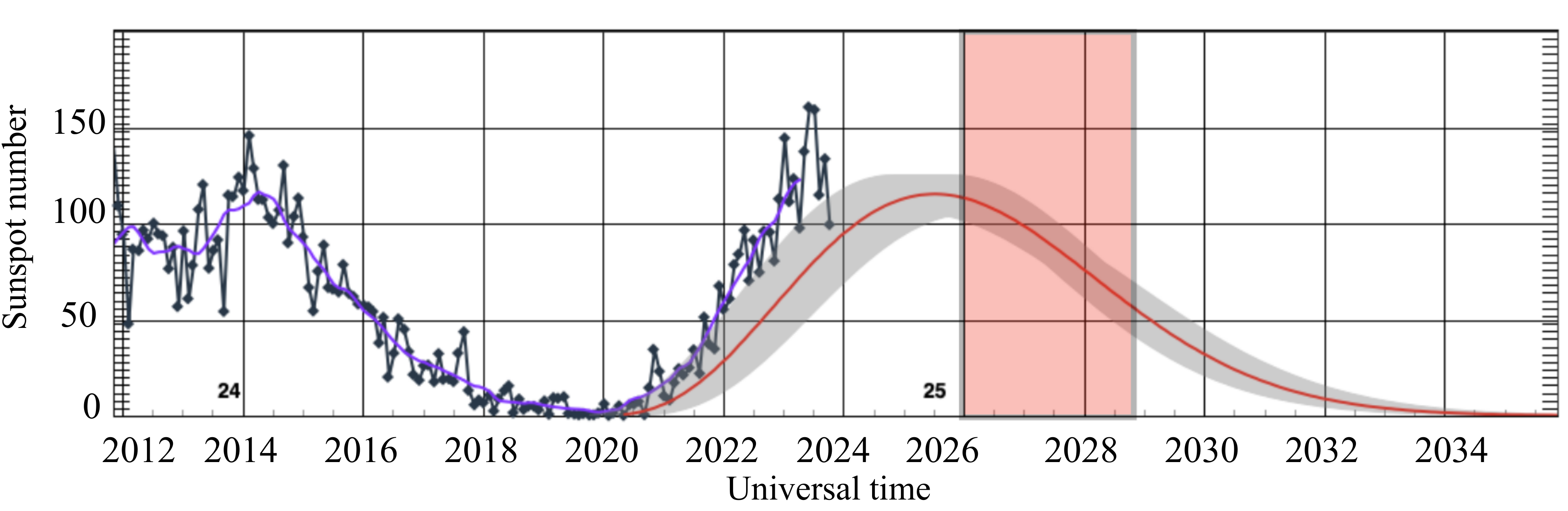}
\caption{The observed (points) and predicted (red line) solar cycle sunspot number from the NOAA website~\cite{NOAA}. The NUSES operation time is highlighted in red. 
\label{fig:solar_cycle}}
\quad
\includegraphics[width=.8\textwidth]{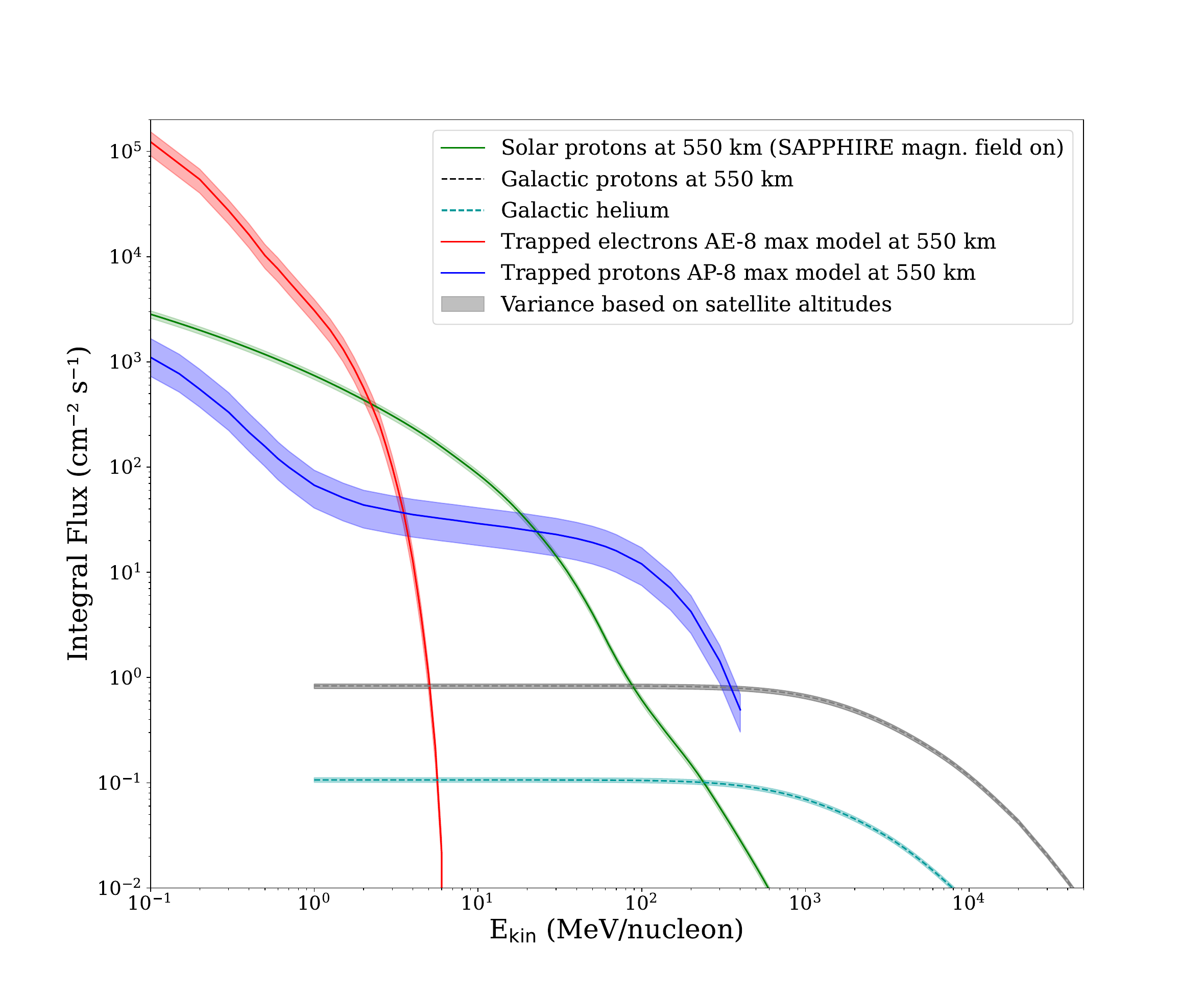}
\caption{Integral fluxes calculated with SPENVIS expected for the NUSES mission of trapped electrons and protons in the Van Allen Belts, solar protons and galactic protons and helium nuclei at 550 km.\label{fig1:fluxes_spenvis}}
\end{figure}

In order to evaluate the radiation damage in space on the sensors of the FPA, we used the SPENVIS (Space Environment Information System)~\cite{SPENVIS} algorithm, a powerful software suite developed by the European Space Agency (ESA). In this framework, we built a project according to the specification of our mission's orbit altitude of 550~km and an inclination of approximately $97.8^{\circ}$ between mid-2026 to mid-2029, which corresponds to an intense solar activity (see figure~\ref{fig:solar_cycle}).

To estimate the total radiation we considered solar particle events (SPEs), trapped protons and electrons in the Van Allen radiation belts and galactic cosmic rays (GCRs). These expected fluxes for the NUSES spacecraft averaged on 15 orbits (1 day) during its operation time are shown in figure~\ref{fig1:fluxes_spenvis}. The shaded areas indicate the fluctuations of these fluxes depending on the variation of the satellite altitude between 500~km and 600~km and considering a period of maximum solar activity. In particular, the higher the altitude, the higher the flux of trapped protons and trapped electrons and consequently the slightly higher the flux of solar protons. Since the flight will occur during the declining part of the solar maximum, the fluxes for the trapped protons and electrons have been calculated using the NASA models AP-8 max and AE-8 max~\cite{vette2,vette1}, respectively, and the solar protons using the magnetic shield on in SAPPHIRE~\cite{sapphire}, and ISO 15390 model~\cite{isomodel} for the GCRs. We used the realistic case of 550~km.
%and therefore not presented in this paper, would be given by an altitude of 600~km by a combination of conditions taking the fluxes for trapped protons and electrons during solar minimum, the flux of solar protons during solar maximum and the flux galactic cosmic rays for the worst solar minimum event in 1996. 
It can be noticed that, for our altitude, the GCR contribution (protons and helium nuclei)\footnote{Heavier nuclei than p GCR fluxes are much lower at these energies and are neglected in this analysis.} is negligible. Moreover, since NUSES will operate during intense solar activity, the solar flux will be very high while the galactic proton flux will be reduced~\cite{osti_644207} as the solar wind impedes GCR penetration into the solar system. The main contributions to the radiation damage to the camera will be due to trapped protons and electrons, and solar protons.  Although it is partially accounted for in the trapped particle models, Earth's albedo has not been evaluated separately because can be considered negligible for a LEO mission.

\section{Dose estimates from radiation in space with Geant4}
\label{sec:geant4}

To estimate the dose that we expect on Terzina's camera, we simulated the geometry of the telescope in Geant4. The simulated mechanics has small differences with respect to the final one defined subsequently, but these few changes are irrelevant for this study. The simulated geometry, similar to the CAD in figure~\ref{fig:FPA_cad}, includes:
\begin{enumerate}
\item an external baffle or tube made of carbon fibre that covers the full telescope;
\item a holder made of carbon fibre, composed of a small baffle and a tower that protects the secondary mirror;
\item different vanes made of carbon fibre;
\item the focal plane composed of the sensors and the PCB connecting the tiles to the rest of the data acquisition system;
\item the optical bench below the PCB (in aluminium and titanium);
\item the aluminium radiators.
\end{enumerate}

We tested different geometrical configurations varying the thickness and height of the external baffle and the thickness of the small baffle, indicated in figure~\ref{fig:FPA_cad}. 
In particular, we report the results for three configurations: the hereafter called \textit{baseline} geometry, with an external baffle of thickness of 1~mm and maximum height of the transverse section of 730~mm and the small baffle of 1~mm thickness; the same external cut baffle and a small baffle 4~mm thick, called \textit{benchmark} configuration;
a better shielded configuration, called \textit{shielded}, consisting of a cylindrical external baffle 2~mm thick and of height 730~mm without a transverse cut, and a small baffle 4~mm thick.

Protons and electrons have been injected into these geometries, simulated in Geant4, to estimate the total radiation dose. Particles were simulated in small energy bins considering the full energy range of the flux calculated with SPENVIS $\frac{d\phi}{dE}$ integrated over a sphere of radius $R=50$~cm surrounding the considered volumes~\footnote{It is important to note that, in this study, the fluxes were considered to be isotropic, which is a conservative case compared to the real scenario. The direction and regions where we expect the presence of these particles have been produced in SPENVIS.}and for the duration of the mission $t_{\mathrm{EoL}}$ of 3 years. The total energy deposited in the sensitive volume of the FPA tiles and in the PCB, $E_{\mathrm{dep}}$ returned in MeV, has been calculated by Geant4~\footnote{Secondary particle production has been simulated and taken into account. Material activation, on the other hand, was considered to be a minor concern and negligible in terms of the total dose.}. 
The total dose released in the corresponding volumes for all particles in the energy bin has been calculated by summing all the differential doses:

\begin{equation}
  \Delta D(E) = \frac{4\pi R^2 t_{EoL}}{2 M_{tot}} c_{\rm MeVJ} E_{dep} \Delta \phi(E)
\label{eq:delta}
\end{equation}

returned in J/kg or Gray, which is the SI unit of ionizing radiation dose defined as the absorption of 1~J of radiation energy per kg of the object. We account for the fact that each particle crosses the spherical volume from the two opposite sides of the sphere by dividing by 2 in eq.~\ref{eq:delta}, assuming that volumes have a uniform density and that $M_{\mathrm{tot}}$ is their total mass (sensors or PCB), and $c_{\mathrm{MeVJ}}$ the conversion factor from MeV to Joule. 
These total doses for a 3-year mission, released in the simulated Terzina geometries by trapped electrons and protons, and solar protons are summarized in table~\ref{tab:dose}.
Considering these results, the shield power and the allowed weight for the mission, the \textbf{benchmark} configuration has been chosen for implementation, as the shielded one would not be affordable for weight limitations.

\begin{table}[t!]
\centering
\begin{tabular}{|c|c|c|c|c|}
\hline
& & Trapped $e$ $\mathcal{D}$ (Gy)& Trapped $p$ $\mathcal{D}$(Gy) &  solar $p$ $\mathcal{D}$ (Gy) \\
\hline
\multirow{2}{*}{\small baseline}& sensors & $14.4\pm 2.5$  &  $0.71\pm 0.28$ & $2.35\pm 0.04$ \\
& PCB & $5.35\pm 0.82$  & $0.58\pm 0.20$ & $1.32\pm0.02$ \\
\hline
\multirow{2}{*}{\small benchmark} & sensors & $10.6\pm 3.3$ & $0.64 \pm 0.23$ & $2.06\pm 0.03$ \\ 
& PCB & $5.70\pm 0.92$  & $0.52\pm 0.18$ & $1.27\pm 0.02$  \\ \hline
\multirow{2}{*}{\small shielded}& sensors & $3.65 \pm  0.52 $&  $0.50\pm 0.17$ & $0.81\pm 0.01$ \\
& PCB & $2.10\pm 0.28$  & $0.49\pm 0.17$ & $0.64\pm 0.01$ \\
\hline
\end{tabular}
\caption{\label{tab:dose} Simulation of the expected dose for a 3 years mission at 550~km from SPENVIS and Geant4. The symmetric uncertainties have been calculated as the root mean square using the estimated doses for different altitudes within 500~km and 600~km.}
\end{table}

\section{Radiation damage in silicon with proton irradiation} %model 3
\label{sec:protons}

\begin{figure}[t!]
    \centering
    \includegraphics[width=1.\textwidth]{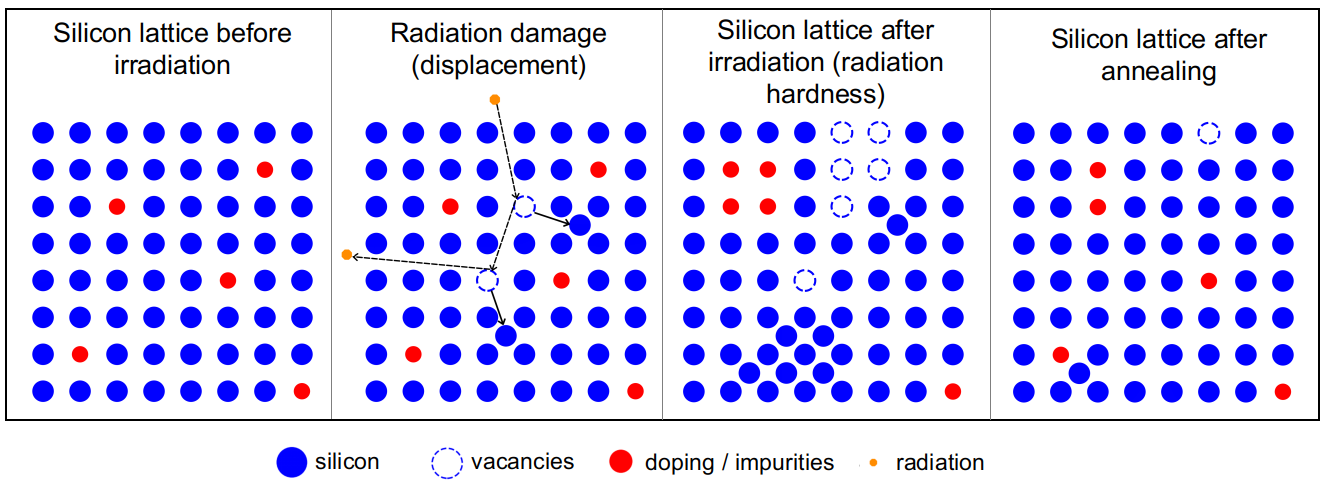}
    \caption{Schemes of the SiPMs lattice at the atomic scale in 2-dimensions to visualize the effect of radiation damage: in the first panel on the left, the lattice has even spacing between doping or impurities; during irradiation (second panel), some single atom displacements occur (the minimum energy for one Si atom displacement is 25~eV). Effects due to higher energy particles forming cluster defects are shown in the third panel. The fourth panel on the right shows the same lattice after the annealing procedure.}
    \label{fig4:Radiation_hardness}
\end{figure}

After the full characterisation of the sensors in section~\ref{sec:terzinaplane}, we have irradiated them with protons and electrons in order to estimate the radiation damage. When an energetic charged particle hits a SiPM, it deposits energy through both ionizing and non-ionizing processes. 
Radiation damage of SiPMs has been extensively discussed in \cite{Garutti:2018hfu}. 
In silicon, bulk damage is due to Non-Ionizing Energy Loss (NIEL) and surface damage by Ionizing Energy Loss (IEL). 
Bulk damage is due to high-energy particles (can be protons, neutrons, electrons, pions or photons), which can displace atoms out of their lattice creating defects as shown in figure~\ref{fig4:Radiation_hardness}. At energies beyond the keV scale, cluster defects can form when more atoms are displaced. Electrons beyond energies of 1~MeV typically can produce single defects, whereas heavier particles like protons also produce cluster defects. In the NIEL hypothesis, the radiation damage is proportional to the non-ionizing energy loss of the particles and this energy loss is proportional to the energy used to dislocate the lattice atoms (displacement energy). 
The defects in silicon crystals lead to an increase in leakage current due to the generation of electron-hole pairs from defects in the depletion region and trap-assisted tunnelling. This leads to a temperature-dependent increase in DCR, which is more significant for protons than for electrons, which primarily induce ionisation rather than defects due to their lower mass and less efficient energy transfer to the crystal lattice, resulting in a minimal increase in DCR.
A detailed computation of the NIEL for protons and electrons in silicon can be done with the Screened Relativistic (SR) Treatment for NIEL Dose Nuclear and Electronic Stopping Power Calculator~\cite{SR}. Hence, we expect that the damage caused by electrons induces a lower increase in DCR when compared to protons at equal over-voltage, temperature, cell sizes, and dose.

\begin{figure}[t!]
    \centering
    \includegraphics[width=0.8\textwidth]{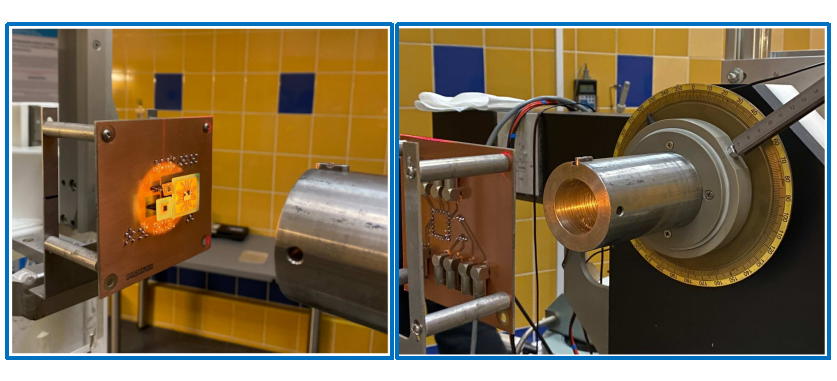}
    \caption{The experimental setup for proton SiPM irradiation at the IFJ PAN in Krakow~\cite{SWAKON20101469}.}
    \label{fig:ExperimentalSetup_proton}
    \end{figure}
    
\begin{figure}[t!]
\centering
\includegraphics[width=0.9\textwidth]{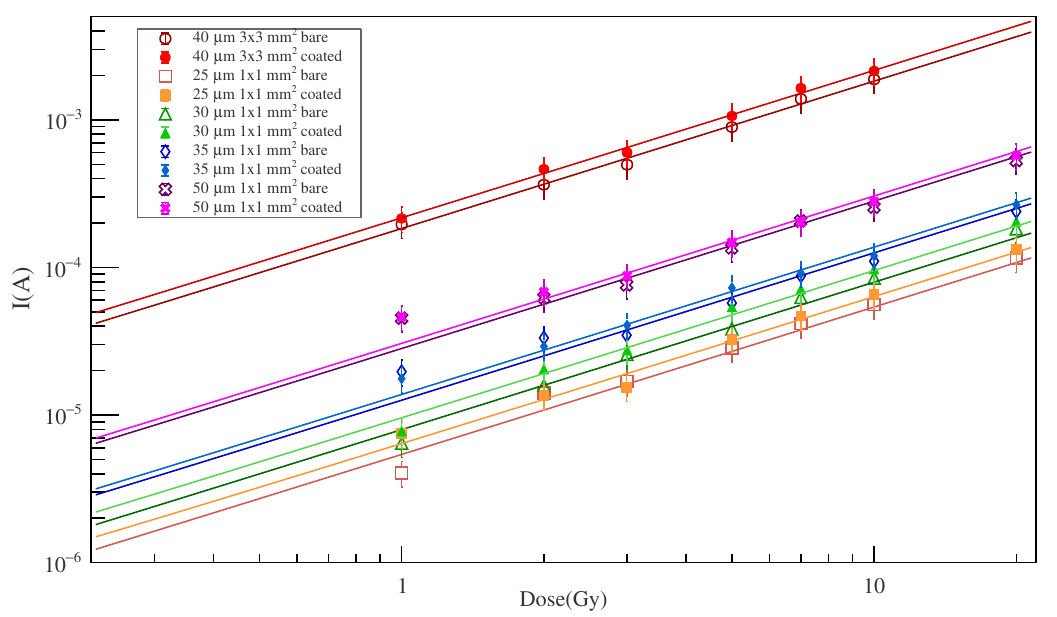}

\caption{Measured currents (symbols with statistical errors) at room temperature at $V_{\mathrm{OV}}= 8.5$~V as a function of the irradiated dose with protons. It is noticeable that the larger the $\mu$-cell size, the larger is the DCR induced by radiation damage. This is another confirmation that 30$~\mu$m is a good choice for Terzina. The lines represent the fit results. 
\label{fig:IV_dose_Fits}}
\end{figure}

We performed the proton irradiation tests at the IFJ PAN in Krakow~\cite{SWAKON20101469} with a 50~MeV proton beam (see the experimental setup in figure~\ref{fig:ExperimentalSetup_proton}). The proton beam spot had a circular shape with a 35~mm diameter and homogeneity better than 5\% with respect to the mean fluence. We measured the currents for $1\times1$~mm$^2$ SiPMs with $\mu$-cell size of 25~$\mu \text{m}$, 30~$\mu \text{m}$, 35~$\mu \text{m}$ and 50~$\mu \text{m}$ and 3$\times$3~mm$^2$ 
SiPMs with a $\mu$-cell size of 40~$\mu \text{m}$ with and without resin coating. We measured the current of the SiPMs as a function of the bias voltage in the post-breakdown Geiger regime. The currents acquired just before the start of and during the irradiation process are shown as a function of the dose for the different devices in figure~\ref{fig:IV_dose_Fits}. The expected linear behaviour has been fitted as: 
\begin{equation}
\label{eq:dose_fit}
I(\mathcal{D})=I_0+m \mathcal{D}
\end{equation}
where the fit parameters, $m$ and $I_0$, are reported in table~\ref{tab:dosefit_postbreakdown} for $V_{\mathrm{bias}}=42$~V. The empirical model obtained from the fit will be used to predict the increase of the DCR as a function of dose in section~\ref{sec:annealing}. Variations of other SiPM features, such as breakdown voltage and quenching resistance, have not been observed after irradiation. 

\begin{table}[htbp]
\centering
\begin{tabular}{|c|c|c|c|c|c|c|}
\hline
sensor size & cell size & $I_0$ (A) & $m$(A/Gy) \\
\hline
 $1\times1$~mm$^2$& 25 $\mu $m coated & $ 6.78\cdot 10^{-8} \pm 1.37\cdot 10^{-8}$ & $6.35\cdot 10^{-6}$\\
  $1\times1$~mm$^2$& 30 $\mu $m coated & $5.77\cdot 10^{-8} \pm 1.17\cdot 10^{-8}$ & $9.53\cdot 10^{-6}$\\
 $1\times1$~mm$^2$& 35 $\mu $m coated & $7.20\cdot 10^{-8} \pm 1.45\cdot 10^{-8}$ & $1.37\cdot 10^{-5}$\\
  $1\times1$~mm$^2$& 50 $\mu $m coated & $1.21\cdot 10^{-7} \pm 2.42\cdot 10^{-8}$ & $ 3.05\cdot 10^{-5}$\\
 \hline
  $1\times1$~mm$^2$& 25 $\mu $m bare & $1.82\cdot 10^{-8} \pm 3.74\cdot 10^{-9}$ & $5.40\cdot 10^{-6}$\\
  $1\times1$~mm$^2$& 30 $\mu $m bare & $ 2.18\cdot 10^{-8} \pm 4.46\cdot 10^{-9}$ & $68.0\cdot 10^{-6}$\\
  $1\times1$~mm$^2$& 35 $\mu $m bare & $ 5.47\cdot 10^{-8} \pm 1.10\cdot 10^{-8}$ & $1.26\cdot 10^{-5}$\\
  $1\times1$~mm$^2$& 50 $\mu $m bare & $1.03\cdot 10^{-7} \pm 2.07\cdot 10^{-8}$ & $2.81\cdot 10^{-5}$\\
\hline
  $3\times3$ mm$^2$& 40 $\mu $m bare & $7.01\cdot 10^{-7}\pm 1.40\cdot 10^{-7}$ & $0.00018287$\\
  $3\times3$ mm$^2$& 40 $\mu $m coated & $7.53\cdot 10^{-7} \pm 1.51\cdot 10^{-7}$ & $0.0002163$\\
\hline
\end{tabular}
\caption{\label{tab:dosefit_postbreakdown} Fit parameters for the current as a function of the dose linear plot in figure~\ref{fig:IV_dose_Fits} for different devices.}
\end{table}

\section{Thermal effect}
\label{sec:thermic}
\begin{figure}[t!]
\centering
\includegraphics[width=0.9\textwidth]{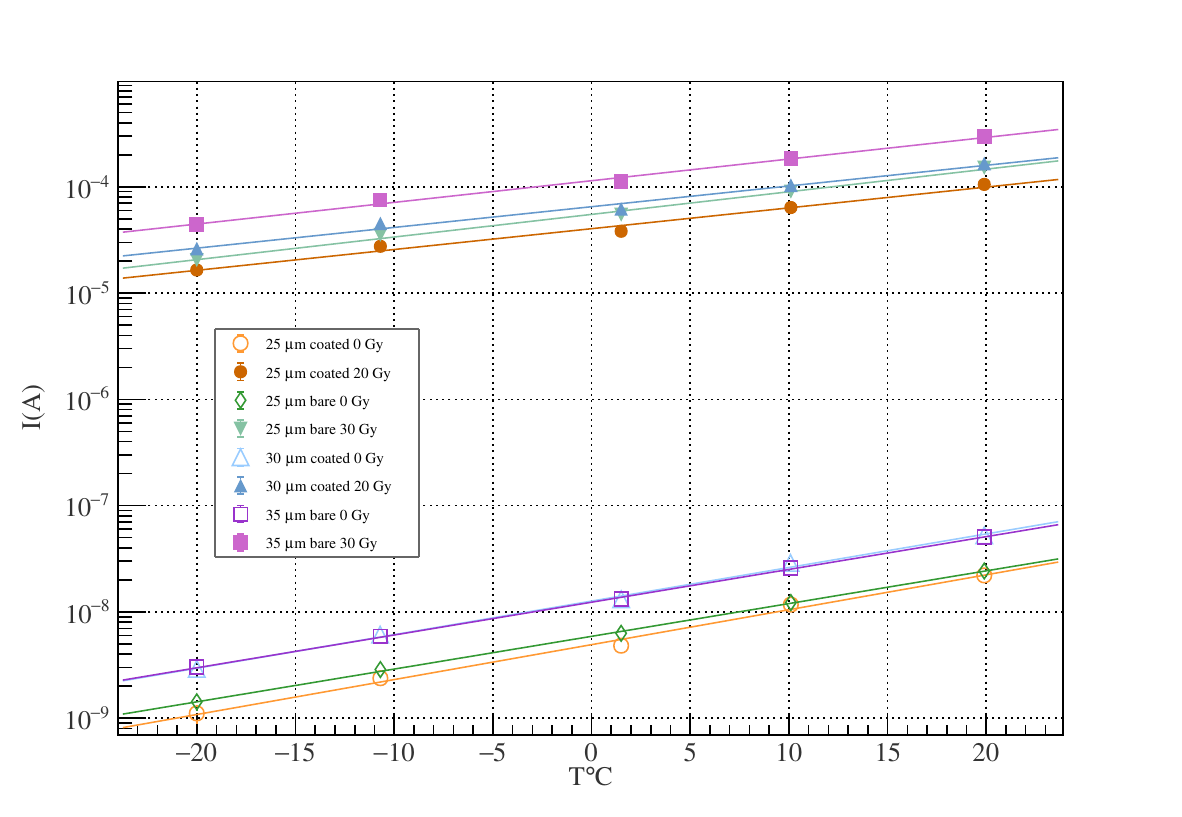}

\caption{Measured currents (symbols with statistical errors) at $V_{\mathrm{OV}}= 8.5$ V as a function of temperature. The lines represent the fit results described by equation~\ref{eq:temperature_correction}.\label{fig:allIT_fits}}
\end{figure}

The optimal operation temperature for Terzina is considered to be -20$^\circ$C and the thermal control will avoid temperatures larger than 5$^\circ$C. In this work, we performed the sensor characterisation at room temperature (see section~\ref{sec:charact}). Nonetheless, we performed some useful IV measurements in a climate chamber to characterise the behaviour of the sensors as a function of their temperature. These allowed us to extrapolate the DCR behaviour at other temperatures than room temperature. Being the DCR proportional to the post-breakdown current at a given over-voltage, we can derive the DCR from the current measurements. The results are used to select the sensors and understand if it was necessary any action to correct the operation temperature. Tests were performed both before and after irradiation. The temperature dependency of the IV curve measurements is modelled with the function, with parameters $I_0$ and $\lambda$:
\begin{equation}
\label{eq:temperature_correction}
I(T)= I_0\cdot e^ {\lambda T}\, .
\end{equation}    
The fit results are reported in table~\ref{tab:lambda_fit} and the measurements and the fit curves are shown in figure~\ref{fig:allIT_fits}. 
It can be noted that the obtained parameter, $I_0$, corresponds to the current at $T (^{\circ}\mathrm{C})=0$ and the rate constant $\lambda$ related to the exponential temperature constant $\tau=1/\lambda$ provides the temperature correction factor used to obtain the currents at the desired temperature throughout this work.
We can estimate the change in current per degree in order to get the temperature correction that we used in all the measurements. For example, we can estimate that, if the current value is doubled, the temperature will be obtained as $2 I_0 = I_0 \cdot e^ {\lambda T_{I_2}}$ as: $T_{I_2}=\frac{\ln{2}}{\lambda}$. The current is doubled every 10 degrees before irradiation and every 15 degrees after irradiation (see last column in table~\ref{tab:lambda_fit}). As shown in figure~\ref{fig:allIT_fits}, the increase in DCR with temperature is slightly slower than in the presence of radiation.

\begin{table}[ht]
%\centering
\begin{tabular}{|c|c|c|c|c|}
\hline
cell size & dose (Gy) & $I_0$ (A) & $\lambda\, (\mathrm{1/C^{\circ}})$ & $\Delta T (^{\circ}\mathrm{C})$\\
& & & & for double-current\\
\hline
 25~$\mu $m coated & 0 & $6.77\cdot 10^{-9}$ & $(7.41\pm 0.35)\cdot 10^{-2}$ & $9.35$ \\
 30~$\mu $m coated & 0 & $1.73\cdot 10^{-8}$ & $(6.96\pm 0.35)\cdot10^{-2}$ & $9.96$\\
 25~$\mu $m bare & 0 & $8.19\cdot 10^{-9}$ & $(6.83\pm 0.40)\cdot10^{-2}$ & $10.15$\\
 30~$\mu $m bare & 0 & $8.41\cdot 10^{-9}$ & $(6.87\pm 0.35)\cdot10^{-2}$ & $10.09$\\
 35~$\mu $m bare & 0 & $1.64\cdot10^{-8}$ & $(6.91 \pm 0.35)\cdot10^{-2}$ & $10.03$\\
\hline

 25 $\mu $m coated & 20 & $5.69\cdot 10^{-5}$ & $(4.55 \pm 0.35)\cdot10^{-2}$  & $15.24$ \\
 30 $\mu $m coated & 20 & $9.093\cdot 10^{-5}$ & $(4.46\pm 0.35)\cdot 10^{-2}$ & $15.54$ \\
 25 $\mu $m bare & 30 & $7.93\cdot 10^{-5}$ & $(4.74\pm 0.35)\cdot10^{-2}$ & $14.61$ \\
 35 $\mu $m bare & 30 & $1.63\cdot10^{-4}$ & $(4.49 \pm 0.35)\cdot10^{-2}$ & $15.44$ \\

\hline
\end{tabular}
\caption{Fit results for the temperature IV model (equation~\ref{eq:temperature_correction}).\label{tab:lambda_fit}}
\end{table}
As in section~\ref{sec:terzinaplane}, we estimated the breakdown voltage using the second log-derivative method. We found that the breakdown voltage changes linearly with temperature by about 30~mV per degree on average. We found a rate for the bare sensor of about 32~mV per degree and a rate of the 28~mV per degree for the coated sensor. On the other hand, the quenching resistance is not affected by the change in temperature.

%All figures and tables should be referenced in the text and should be
%placed on the page where they are first cited or in
%subsequent pages. Positioning them in the source file
%after the paragraph where you first reference them usually yields good
%results. See figure~\ref{fig:i} and table~\ref{tab:i} for layout examples. 
%Please note that a caption is mandatory and it must be placed at the bottom of both figures and tables.

\section{Electrons irradiation measurements and simulations}
\label{sec:electrons}

\begin{figure}[t!]
		\centering
            \color{red} % Set the color to red
            \fbox{\includegraphics[height=0.2\textheight,width=0.65\textwidth]{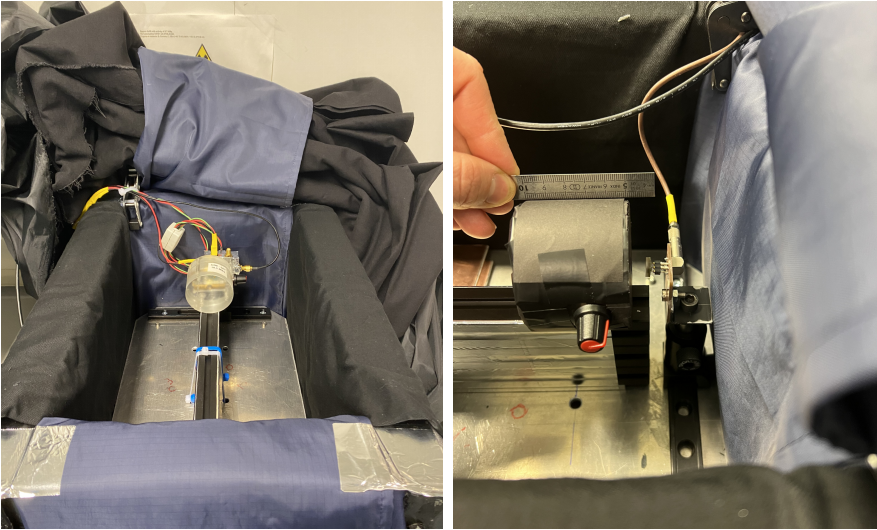}}
			
	\caption{ The experimental setup for irradiation of SiPMs with a $^{90}Sr$ source. }
	\label{fig6.2:ExperimentalSetup}
	\end{figure}

\begin{figure}[t!]
\centering
\includegraphics[width=\textwidth]{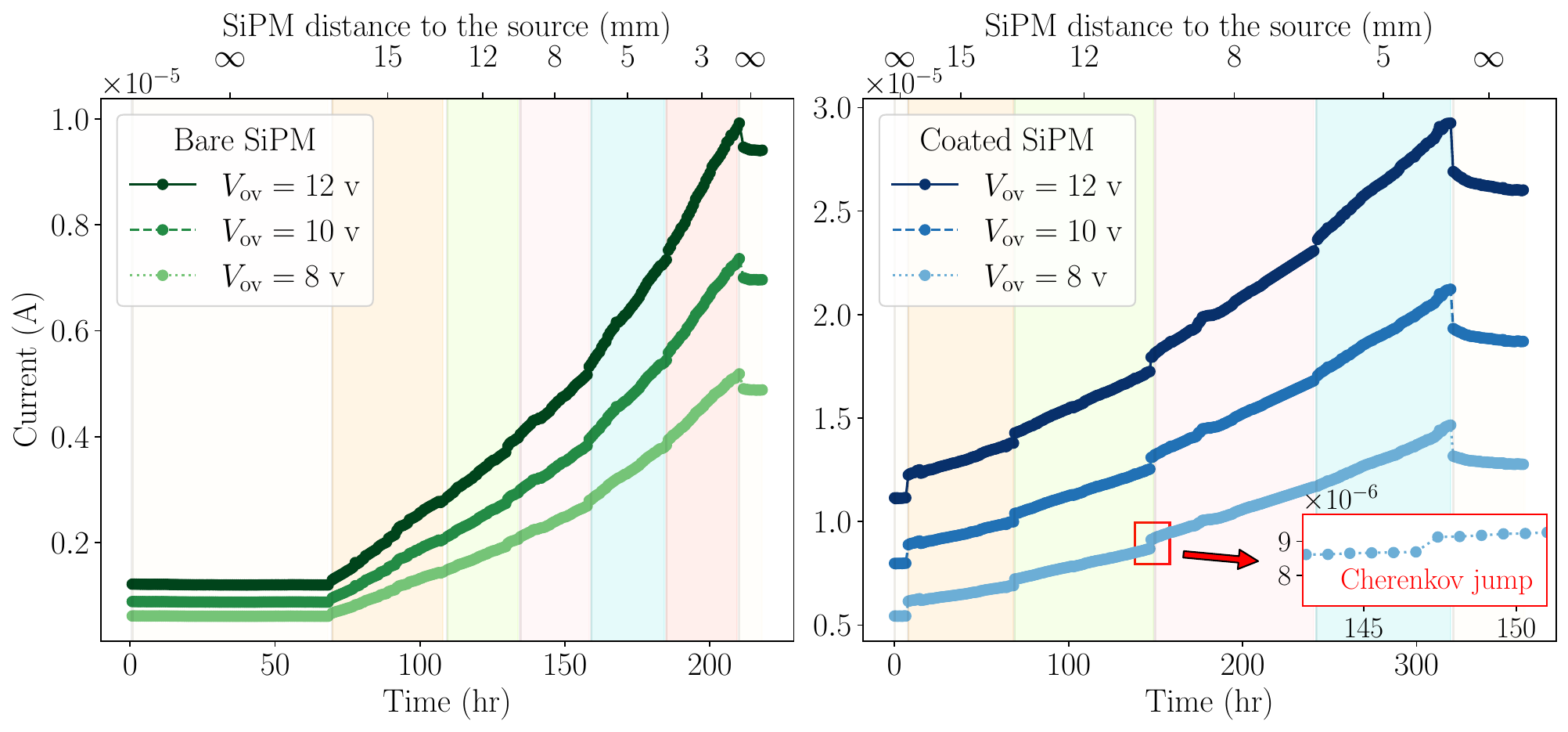}
\caption{The current as a function of time for three over-voltage values after temperature correction (using equation~\ref{eq:temperature_correction}), for bare SiPMs (left) and coated ones (right). Each indicated time range, highlighted with different colours and on the top horizontal axis, corresponds to the indicated distance between the source and SiPM starting from no source ($\infty$). The zoomed plot in the insert on the right panel for the coated SiPM shows one of the resulting Cherenkov jumps due to moving the source closer to the SiPM.}
\label{fig6.5:ItCorrection}
\end{figure}

In order to measure the radiation damage effect on SiPMs, namely the increase of DCR due to electrons, and predict this effect during the NUSES mission, we used a strontium~90 ($^{90}Sr$) $\beta$ emitting source with activity of 16.75~MBq. Two bare and coated (resin 0.75~mm thick) NUV-HD-MT of $3 \times3$~mm$^2$ SiPMs with 40~$\mu\text{m}$-cell size and $8.52\,\text{mm}^2$ sensitive area were exposed to the source. All measurements are done in an over-voltage interval from 0 to 18~V. Considering the energy spectrum of primary and secondary electrons from $^{90}Sr$ decays~\cite{osti_5143758}, with this source, we can study radiation damage from electrons up to a maximum energy of 2.28~MeV. 
As done for protons, we measured the current increase, but in this case, we moved the source at different distances from the SiPM using the experimental setup shown in figure~\ref{fig6.2:ExperimentalSetup}. We also built up a full simulation in Geant4 of this setup to compare with the measurements. Figure~\ref{fig6.5:ItCorrection} depicts the measured current as a function of time at different over-voltages and changing the distance of the source from the SiPM as indicated on the top axis. 

\begin{figure}[t!]
		\centering
		\includegraphics[width=1\textwidth]{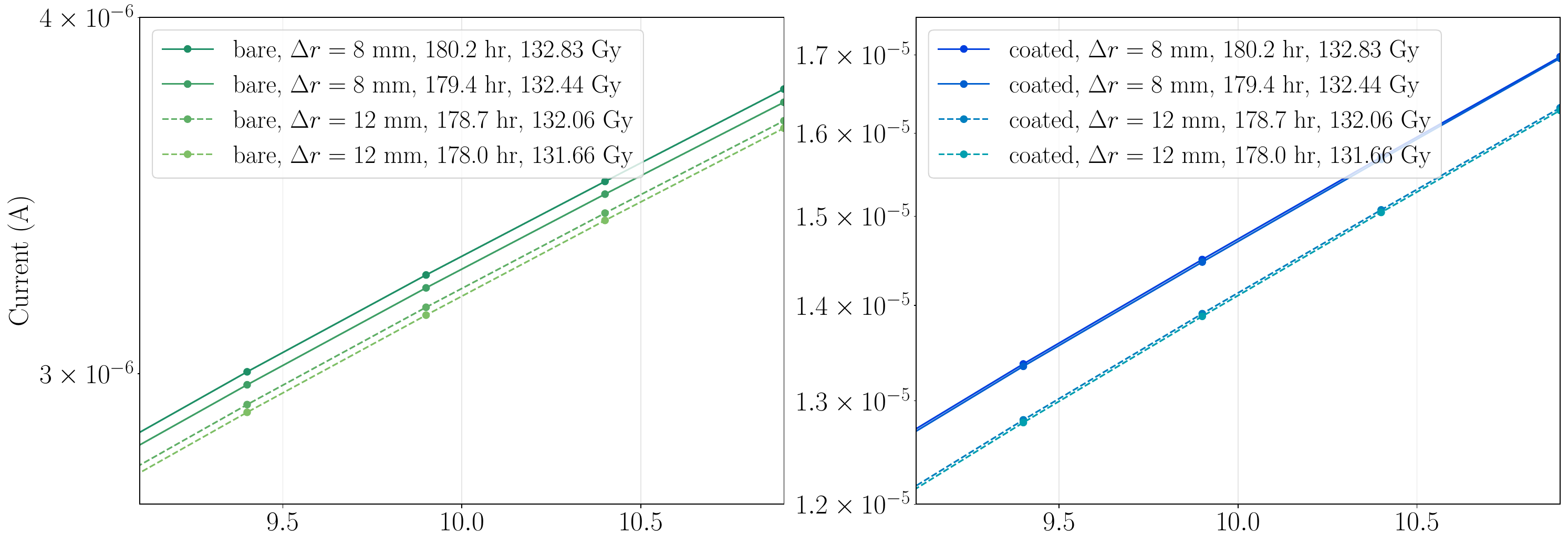}
		\caption{The IV curve for the same bare SiPM (left) and coated SiPM (right), measured at different times. The lower two curves are measured with the source at a 12~mm distance from the SiPM ($\Delta r = 12$~mm), while the other two were taken with the source positioned 8~mm away from the SiPM ($\Delta r = 8$~mm). The corresponding dose is included in the plot by using the results from the simulation and equation~\ref{Dose_calc}.}
		\label{fig6.3:IVCurveSample}
	\end{figure}

Figure~\ref{fig6.3:IVCurveSample} shows the same current as a function of over-voltage (IV curves) for the same sensors, with bare SiPM on the left panel and coated SiPM on the right panel. The time of irradiation for each IV curve is indicated on the label. First, we put the source at a 12~mm distance from the SiPM (lower dashed curves). It can be seen that the different times of exposure to the source introduce a small difference due to the electron radiation damage for different exposure times. Then we moved the source to 8~mm distance to the SiPM and we again see that for different exposure times the effect on the IV curve is small between, but on the right panel for the coated SiPM there is a larger increase as we moved the source closer to the SiPM. This indicates that, other than radiation damage, there is an additional contribution to the current increase, which is due to the Cherenkov emission of electrons in the coating resin. As expected, this effect is not observed in the left panel for the bare SiPM.

The indicated doses in figure~\ref{fig6.3:IVCurveSample} are calculated as follows.
The simulation provides us with the precise energy deposition ($dE/dx$) of electrons resulting from ionizing and non-ionizing processes as they pass through each layer of the SiPM. By dividing the silicon volume into 8 layers (with the thickness of 60~$\mu$m for each layer) and running the simulation many times, we obtained the mean value of the total energy deposit in the volume of the SiPM, denoted as $dE/dx|_{\text{mean}}$. We can then compare the $dE/dx|_{\text{mean}}$ or dose in each layer to analyse where most of the OCT and DCR contributed. Using the following formula, we calculate the dose ($\mathcal{D}$) in each layer:
	
\begin{equation}
\label{Dose_calc}
\mathcal{D} = \frac{dE/dx|_{\text{mean}} \; . \; e \; . \; \mathcal{A}}{m_{\text{V}}}
\end{equation}
	
\begin{figure}[t!]
		\centering
		\includegraphics[width=\textwidth]{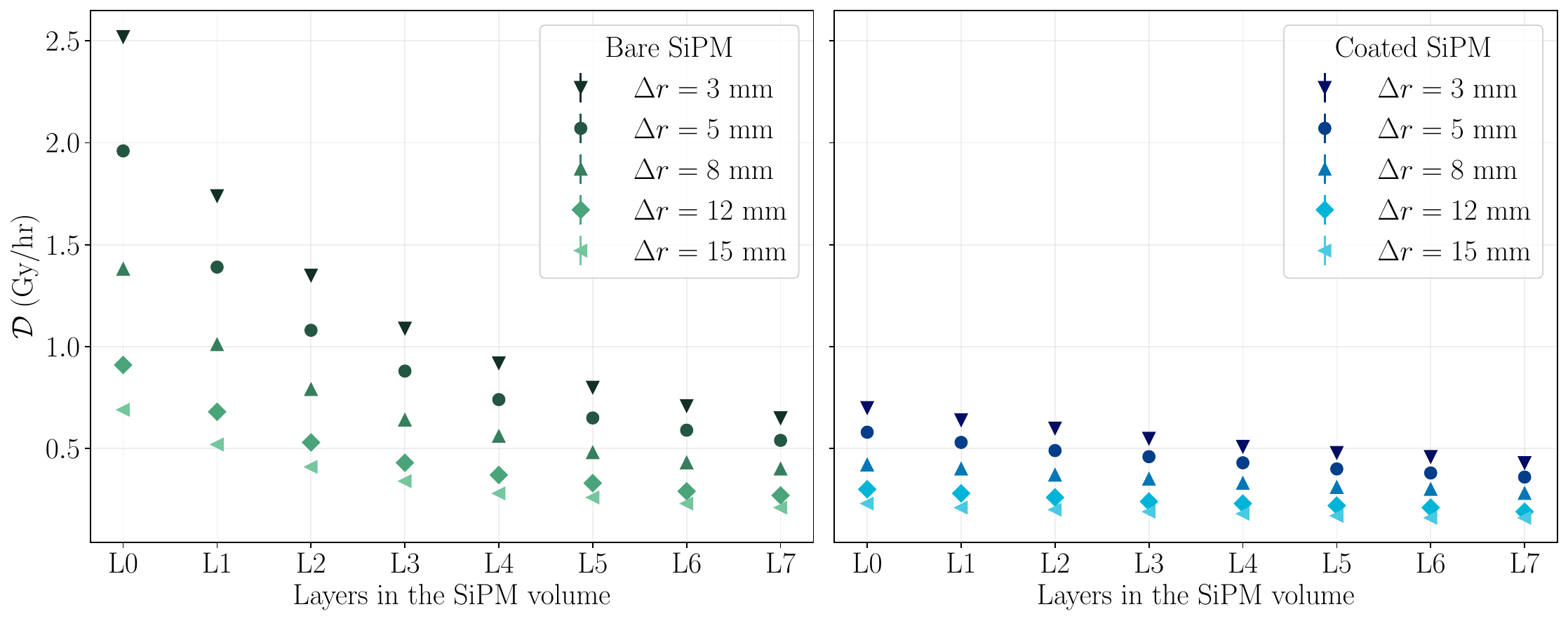}
		\caption{Dose in Gy/hr at each layer (L) from simulation of a bare (left) and coated (right) SiPM for various distances between SiPM and $^{90}Sr$ ($\Delta$r).}
		\label{fig:DoseInLayer}
	\end{figure}

Here, $e$, $\mathcal{A}$, and $m_{\text{V}}$ represent the charge of an electron, the activity of the source, and the mass of the silicon layer, respectively. It is worth noting that each simulation iteration corresponds to a single decay event of $^{90}Sr$. Considering the source's activity, this equation provides the equivalent of a $dE/dx|_{\text{mean}}$ per one second, and thus, we derive the dose rate in Gy/hr. Figure~\ref{fig:DoseInLayer} shows the final dose from the simulation in each silicon layer of the SiPM. The upper layers absorb the highest dose and thus experience the most radiation damage. Additionally, comparing the bare and coated SiPM configurations reveals that the 0.75~mm-thick resin on top of the SiPM reduces the dose received by the layers of the SiPM.

Overall, we use the final estimated dose as a function of time from simulation to estimate the irradiated dose at the time of IV measurements in figure~\ref{fig6.3:IVCurveSample}. This will give us the expected current increase per received dose for the trapped electrons.

%We discourage the use of inline figures (e.g. \texttt{wrapfigure}), as they may be
%difficult to position if the page layout changes.

%We suggest not to abbreviate: ``section'', ``appendix'', ``figure''
%and ``table'', but ``eq.'' and ``ref.'' are welcome. Also, please do
%not use \texttt{\textbackslash emph} or \texttt{\textbackslash it} for
%latin abbreviaitons: i.e., et al., e.g., vs., etc.

\section{DCR prediction during mission}
\label{sec:9}

The DCR represents the baseline count rate in the absence of incident light. In order to avoid that the camera is constantly triggering on DCR induced events, the threshold of the trigger has to be increased with increasing DCR. This impacts the lowest achievable energy of the detected primary cosmic rays. With all the measurements that have been done for thermal response, electrons and protons campaigns, we can now calculate the DCR at a certain operation temperature (e.g. at $T_{\mathrm{op}}=0^\circ$C) considering the overall effects through the following formula:
\begin{equation}
    \label{DCR}
    \text{DCR} (\text{$\mathcal{D}$}|\textrm{V$_{\textrm{ov}}$}, T_\mathrm{op}) \propto I^\ast(\textrm{V$_{\textrm{ov}}$}, \text{$\mathcal{D}$},T) \times e^{\lambda (T_{\mathrm{op}}-T)} \times \frac{1}{q\,\text{Gain}} \times \mathcal{S}
\end{equation}
Here, the $V_{\textrm{ov}}$, $\mathcal{S}$, Gain, $T_\mathrm{op}$, $\mathcal{D}$, and $q$ stand for over-voltage ($V_{\textrm{ov}}=10$ V), effective sensitive area, gain of the SiPM in use, average operation temperature during the mission, dose released by either protons or electrons, and the electron charge, respectively. 

\begin{figure}[t!]
    \centering
    \includegraphics[width=1.\textwidth]{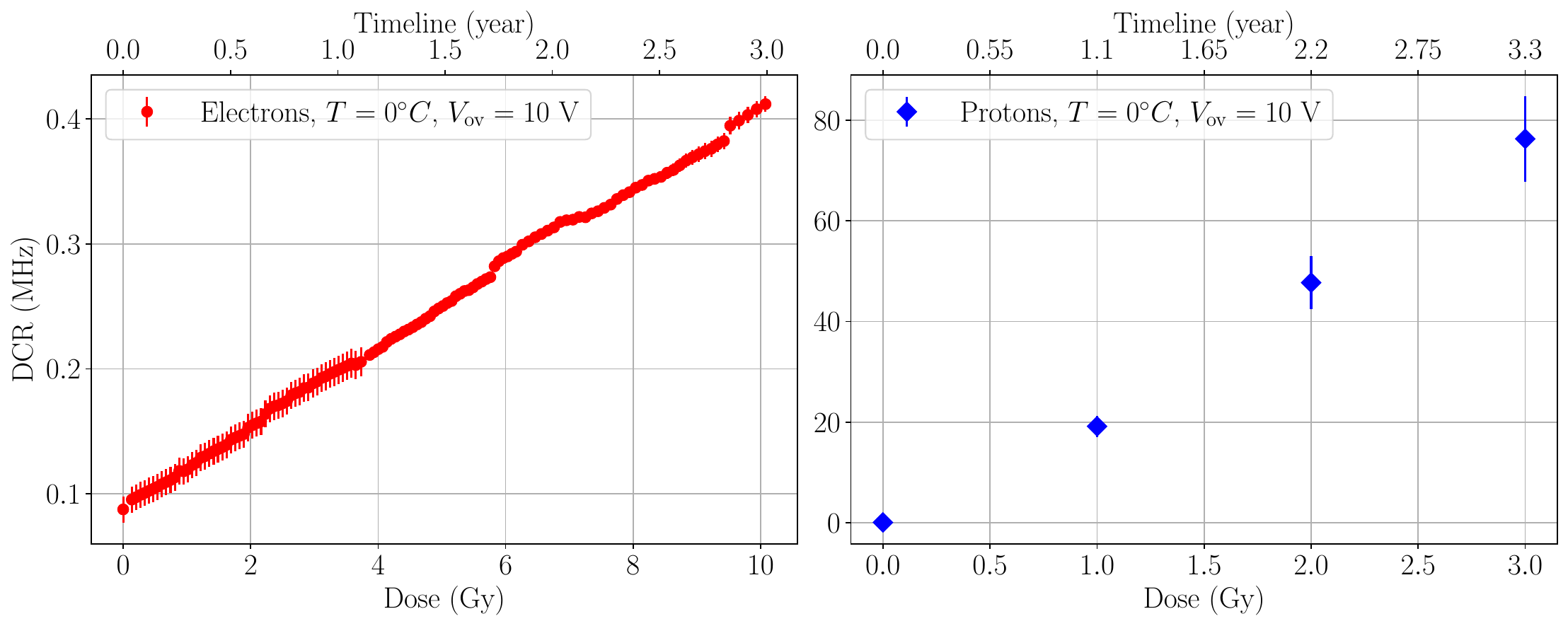}
    \caption{Inferred DCR per pixel of Terzina SiPMs with sensitive area of $\sim 6.58~\text{mm}^2$ from IV measurements during the irradiation as a function of dose and mission timeline. The errors are due to the instrument scale precision, which changes with range as can be seen in the plot.}
    \label{fig5:DCR_proton_electron}
\end{figure}

The $ I^\ast(\textrm{V$_{\textrm{ov}}$}, \text{$\mathcal{D}$},T)= \frac{I(\textrm{V$_{\textrm{ov}}$}, \text{$\mathcal{D}$}, T)}{s^\ast}$ corresponds to the measured current $I(\textrm{V$_{\textrm{ov}}$}, \text{$\mathcal{D}$},T)$ divided by $s^\ast$, namely scaling for the ratio of areas to take into account the different SiPM sizes. The exponential term provides the measured current at the desired temperature following the model in equation~\ref{eq:temperature_correction}. Combining the DCR results with the dose-per-time response from the simulation yields the results in figure~\ref{fig5:DCR_proton_electron}.
As can be seen from the graph, the damage caused by electrons is very low when compared to protons at equal over-voltage, temperature, cell sizes, and dose, as expected. Specifically, 2~Gy of electrons corresponds to a DCR of approximately 0.15~MHz, while 2~Gy from protons results in a DCR of around 50~MHz, starting from the same DCR at 0~Gy. 

 For this reason, in the next section, only the radiation damage produced by protons will be taken into account in the thermal annealing strategy done to recover the SiPM response.
 
\subsection{Annealing}
\label{sec:annealing}

A mitigation strategy based on thermal annealing is proposed for addressing radiation damage of the sensors which causes DCR increase, then resulting in higher power consumption and higher energy threshold. This is a controlled thermal process used to improve the performance and stability of electronic devices, including SiPMs. During annealing, SiPMs are subject to elevated temperatures for a specific period, followed by gradual cooling. This process helps to repair structural defects and radiation-induced damage that affect the device's performance. Additionally, annealing can reduce internal stresses in the semiconductor material, thereby improving the uniformity of the SiPM's response and reducing background noise.

\begin{figure}[t!]
\centering
\includegraphics[width=0.9\textwidth]{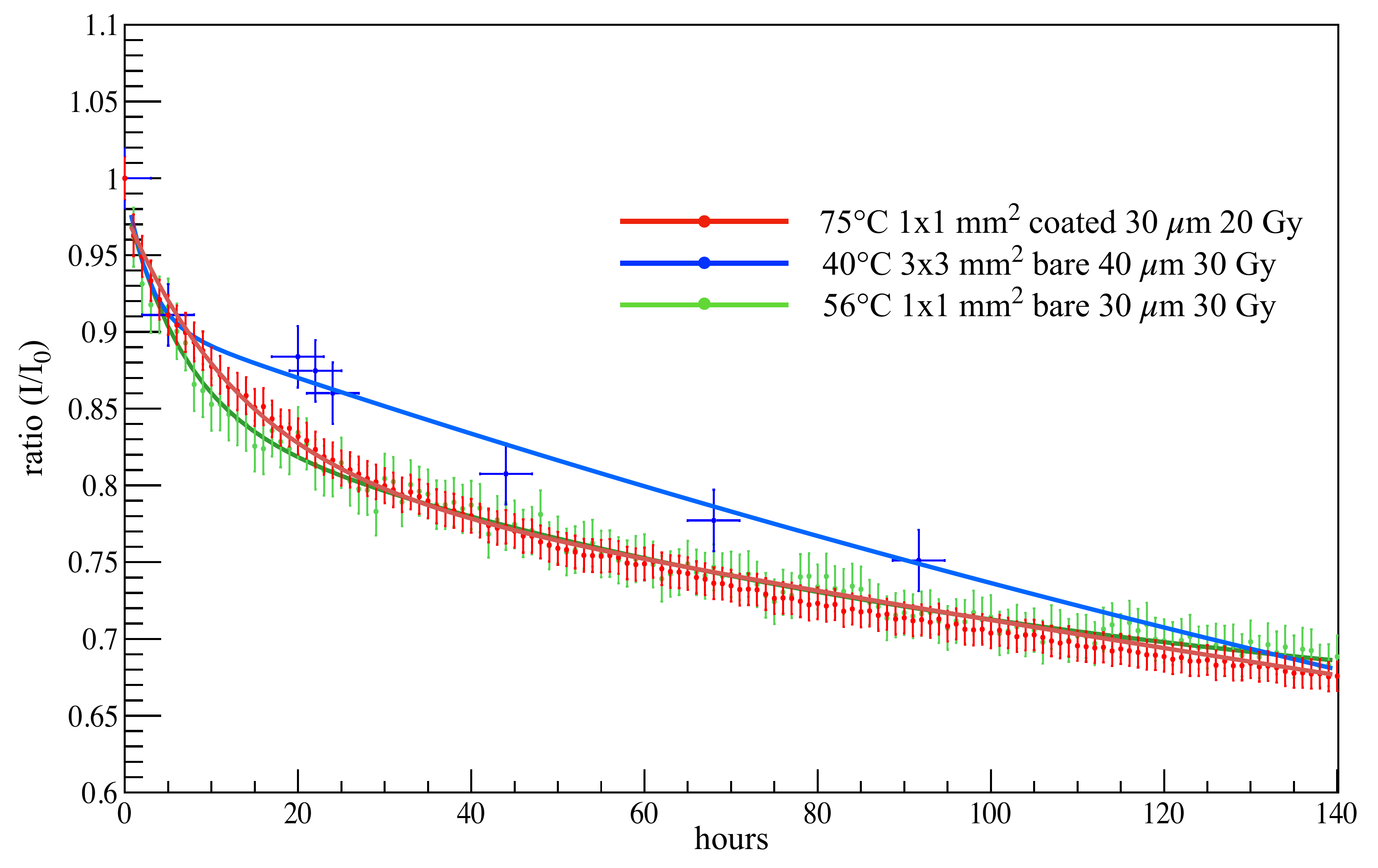}
\caption{Normalised current measured at 10 V over-voltage vs the time after the SiPM irradiation when they are stored at different temperatures (in the legend). The curves are fitted with an exponential function, for different sensors.\label{fig:IV_time}}
\end{figure}

In the laboratory, we studied the annealing effect by placing the irradiated devices in a climate chamber at a fixed temperature for a certain period and regularly measuring their IV curve. Then we measured again the current and compare it with the SiPM current before the annealing at the given over-voltage of 10~V to determine the recovery efficiency of the annealing process. 
The time evolution of the 10~V over-voltage current for different storage temperatures is shown in figure~\ref{fig:IV_time} for three types of SiPMs. In this figure, the y-axis shows the normalised current, meaning the dark current relative to that measured during the first measurement after irradiation. 
It can be noticed that the current is following an exponential decay with time, whose amplitude, time constant, and offset vary with the storage temperature. At time $t_0 = 0$ h, the measurement taken immediately after irradiation is shown (normalised current is equal to 1). In the plot, the first annealing measurement corresponds to time $t_1> t_0$. This takes into account the effect of the sensors being subjected to annealing without heating during transport from the irradiation site. The data points in figure~\ref{fig:IV_time} are therefore fitted with the following function:

\begin{equation}
R= c_1 \exp{-t/s_1} + q + c_2 \exp{s_2 (t+1)}
\label{eq:exp}
\end{equation}

whose fit parameters are reported in table~\ref{tab:annealing_fit}. The exponential offset is linked to the fraction of the post-irradiation dark current that can be recovered after an infinite amount of time, while the exponential slope can be seen as a dark current recovery rate. Based on various measurements for different SiPMs at fixed temperatures, we concluded that at temperatures above $50^\circ \text{C}$ and an annealing cycle of 84~hours, up to $40\%$ of the SiPM's response current can be recovered.

\begin{table}[htbp]
\centering
\begin{tabular}{|c|c|c|c|c|c|c|}
\hline
T ($^{\circ}$C) & sensor &  $\mathcal{D}$ (Gy) & $c_1$ & $c_2$ & $s_1$ (hr$^{-1}$) & $s_2$ (hr$^{-1}$) \\
\hline

40 & 40 $\mu $m bare 3x3 mm$^2$ & 30 & $0.7$ & $0.13$ & $355$ & $-0.39$\\
56 & 30 $\mu $m bare 1x1 mm$^2$ & 20 & $0.22$ & $0.155$ & $100$ & $-0.153$\\
75 & 30 $\mu $m coated 1x1 mm$^2$ & 20 & $0.7$ & $0.178$ & $650$ & $-0.075$\\

\hline
\end{tabular}
\caption{Annealing exponential model.\label{tab:annealing_fit}}
\end{table}

Using this information, we constructed a model to estimate the expected DCR in space due to accumulating irradiation and the planned annealing processes. These two components change the DCR during the mission and
can be described by the following differential equation:

\begin{equation}
    \frac{d \mathrm{DCR}}{dt}= C_{\mathrm{ir}} + C_{\mathrm{an}}\mathrm{DCR}
\end{equation}
where $C_{\mathrm{ir}}$ and $C_{\mathrm{an}}$ are numeric constants and the first and the second terms represent the irradiation and annealing contributions, respectively. 

\begin{figure}[t!]
    \centering
    \includegraphics[width=0.8\textwidth]{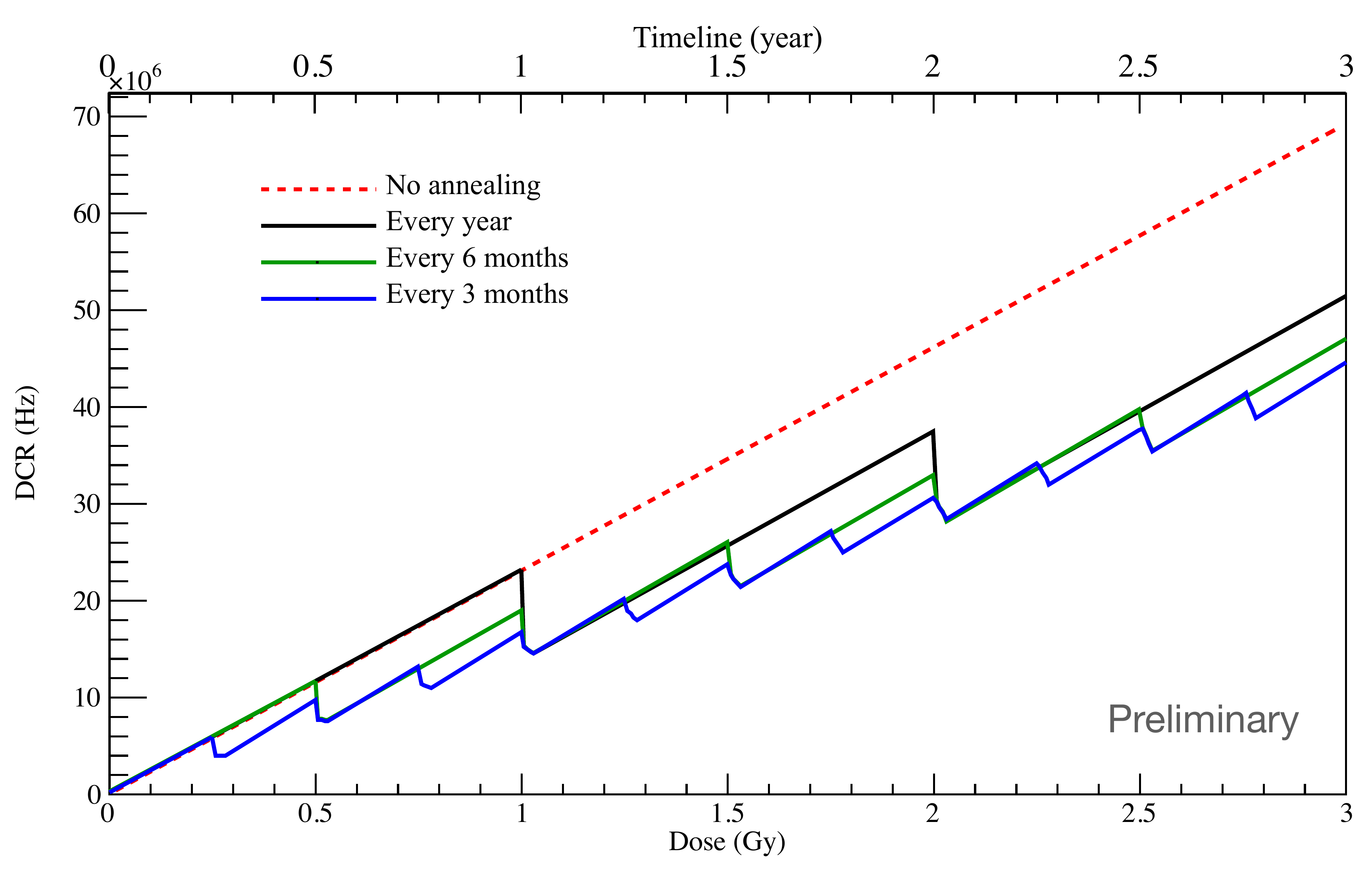}
    \caption{The DCR change from the proton irradiation and annealing cycles during the satellite lifetime.}
    \label{fig6:annealing}
\end{figure}

    In absence of annealing, irradiation scenario, the solution of this equation is given by the linear increase in DCR as a function of the dose described in equation~\ref{eq:dose_fit}. Absent annealing can be defined as the equation being solved at a temperature for which self-annealing is negligible. This is evidenced by the absence of annealing observed in our experimental setup for temperatures below a few degrees.
During the annealing process, the solution of the equation is given by the exponential decay model described in equation~\ref{eq:exp}. Note that over time, the effectiveness of annealing cycles decreases as the lattice becomes more rigid and less responsive to the heating treatment. Moreover, during the annealing cycle the FPA is still taking irradiation. Taking into account all these factors, the resulting DCR from proton irradiation as a function of time/dose for three different annealing strategies is shown in figure~\ref{fig6:annealing}.

\section{Conclusions}
This work presents the characterisation of FBK SiPMs, which allowed to select the optimal size of the $\mu$-cells of SiPMs to adopt in the Terzina mission, and the study that determines the power consumption increase with radiation damage. It also leads to the definition of a mitigation strategy based on periodic annealing during the mission.
A model has been developed from the measurements which allows other users to determine similar mitigation strategies to this relevant damaging effect of SiPMs.

\section*{Acknowledgments}
NUSES is funded by the Italian Government (CIPE n. 20/2019), by the Italian Ministry of Economic Development (MISE reg. CC n. 769/2020), by the Italian Space Agency (CDA ASI n. 15/2022), 
by the European Union NextGenerationEU under the MUR National Innovation Ecosystem grant ECS00000041 - VITALITY - CUP D13C21000430001 and by the Swiss National Foundation (SNF grant n. 178918). 
%This study was carried out also in collaboration with the Space It Up project funded by the Italian Space Agency, ASI, and the Ministry of University and Research, MUR, under contract n. 2024-5-E.0 - CUP n. I53D24000060005. 
The authors would like to thank Officina Stellare S.p.A, in particular Andrea Turella, Andrea Cecconello, Alessandro Scudeler, and Giuseppe Crescenzio, for contributing to the design, manufacturing and assembly of the Terzina Telescope.  

%\section{Conclusions}

%\section{Appendix}

%\begin{figure}[htbp]
%\centering
%\includegraphics[width=.4\textwidth]{example-image-a}
%\qquad
%\includegraphics[width=.4\textwidth]{example-image-b}
%\caption{Always give a caption.\label{fig:i}}
%\end{figure}

% Bibliography

%% [A] Recommended: using JHEP.bst file
\bibliographystyle{JHEP}
%The Bibliography can be built by using biblio.bib in the zip folder or by using BibTex
%\bibliography{biblio.bib}

%BibTex
\providecommand{\href}[2]{#2}\begingroup\raggedright\endgroup
\end{document}